\documentclass[aps,prb,amsmath,amssymb,reprint,preprintnumbers,showpacs,showkeys,intlimits,longbibliography]{revtex4-1}
\usepackage[german,english]{babel}

\bibpunct{[}{]}{,}{n}{}{}

\usepackage[dvipsnames]{xcolor}
\usepackage{bm,latexsym,mathrsfs,enumerate,xcolor}
\usepackage[mathcal]{euscript}
\usepackage[breaklinks=true,unicode=true,urlcolor = blue,colorlinks = true,citecolor = blue,linkcolor = blue]{hyperref}
\usepackage{mathtools}

\usepackage{graphicx}
\DeclareMathOperator{\sech}{sech}
\graphicspath{{./figs/}}
\renewcommand{\vec}[1]{\bm{#1}}
\begin{document}

\title{Localization of magnon modes in a curved magnetic nanowire}

\author{Yuri~Gaididei}
\email{ybg@bitp.kiev.ua}
\affiliation{Bogolyubov Institute for Theoretical Physics, 03143 Kyiv, Ukraine}

\author{Volodymyr~P.~Kravchuk}
\email{v.kravchuk@ifw-dresden.de}
\affiliation{Bogolyubov Institute for Theoretical Physics, 03143 Kyiv, Ukraine}
\affiliation{Leibniz-Institut f{\"u}r Festk{\"o}rper- und Werkstoffforschung, IFW Dresden, D-01171 Dresden, Germany}

\author{Franz~G.~Mertens}
\email{franzgmertens@gmail.com}
\affiliation{Physics Institute, University of Bayreuth, 95440 Bayreuth, Germany}

\author{Oleksandr~V.~Pylypovskyi}
\email{engraver@knu.ua}
\affiliation{Taras Shevchenko National University of Kyiv, 01601 Kyiv, Ukraine}

\author{Avadh~Saxena}
\email{avadh@lanl.gov}
\affiliation{Theoretical Division, Los Alamos National Laboratory, Los Alamos, New Mexico 87545, USA}

\author{Denis~D.~Sheka}
\email[Corresponding author: ]{sheka@knu.ua}
\affiliation{Taras Shevchenko National University of Kyiv, 01601 Kyiv, Ukraine}

\author{Oleksii~M.~Volkov}
\email{o.volkov@hzdr.de}
\affiliation{Helmholtz-Zentrum Dresden-Rossendorf e. V., Institute of Ion Beam Physics and Materials Research, 01328 Dresden, Germany}	

\date{\textcolor[rgb]{0.00,0.50,0.75}{May 28, 2018}}

%

\begin{abstract}
Spin waves in magnetic nanowires can be bound by a local bending of the wire. The eigenfrequency of a truly local magnon mode is determined by the curvature: a general analytical expression is established for any infinitesimally weak localized curvature of the wire. The interaction of the local mode with spin waves, propagating through the bend, results in scattering features, which is well confirmed by spin-lattice simulations.
\end{abstract}




	
\maketitle

\section{Introduction}
\label{sec:intro}

Spin waves as collective excitations of magnetically ordered medium were introduced by \citet{Bloch30} in 1930, where he predicted that spin-waves should behave as weakly interacting quasiparticles obeying the Bose-Einstein statistics. These quasiparticles, as quanta of spin waves, are called magnons: the dynamic eigen-excitations of a magnetically ordered body. A wide variety of linear and nonlinear spin wave phenomena boosted the interest into the fundamental properties of the spin waves~\cite{Holstein40,Dyson56,Akhiezer68}, whereas their transport abilities were of great interest for applications in telecommunication systems~\cite{Gurevich00}. Due to the possibility of building low-power logical devices, spin waves are considered as potential data carriers for computing devices~\cite{Kruglyak10a,Lenk11,Chumak14,Chumak15}. However, in order to exploit spin waves for data processing in real devices, it requires a means to guide spin waves without disturbance in structures with complicated geometry, which in turn requires bent and curvilinear parts to save space on the chips~\cite{Vogt12,Xing13,Haldar16}.

During the past few years, there has been a growing interest in the curvature effects in the physics of nanomagnets. A crucial aspect of the interest is caused by recent achievements in nanotechnologies of flexible~\cite{Perez15}, stretchable~\cite{Melzer11}, and printable magnetoelectronics~\cite{Makarov13b}. 
The topic of magnetism in curved geometries brings about a series of fascinating
geometry-induced effects in the magnetic properties of materials \cite{Streubel16a}.
A fully three-dimensional (3D) approach was put forth recently to study dynamical and static properties of arbitrary curved magnetic shells and wires~\cite{Gaididei14,Sheka15}, and also stripes \cite{Gaididei17a}: the geometrically-broken symmetry in curvilinear magnetic systems results in the emergence of a curvature-induced anisotropy and a Dzyaloshinskii--Moriya interaction (DMI) driven by exchange interaction. The latter effect results in possible magnetochiral effects \cite{Hertel13a}, for a review see Ref.~\cite{Streubel16a}. It is known that the curvilinear-geometry-induced DMI is a source of pinning of domain walls by a local bending of the wire \cite{Yershov15b}, it causes a coupling of chiralities in spin and physical spaces for the M\"{o}bius ribbons \cite{Pylypovskyi15b,Gaididei17a}, it results in a number of nonreciprocal effects in helical wires~\cite{Sheka15c,Yershov16,Pylypovskyi16,Volkov18}, it can even stabilize a skyrmion in a curved magnet \cite{Kravchuk16a,Kravchuk18a}.

\begin{figure*}
	\begin{center}
		\includegraphics[width=\linewidth]{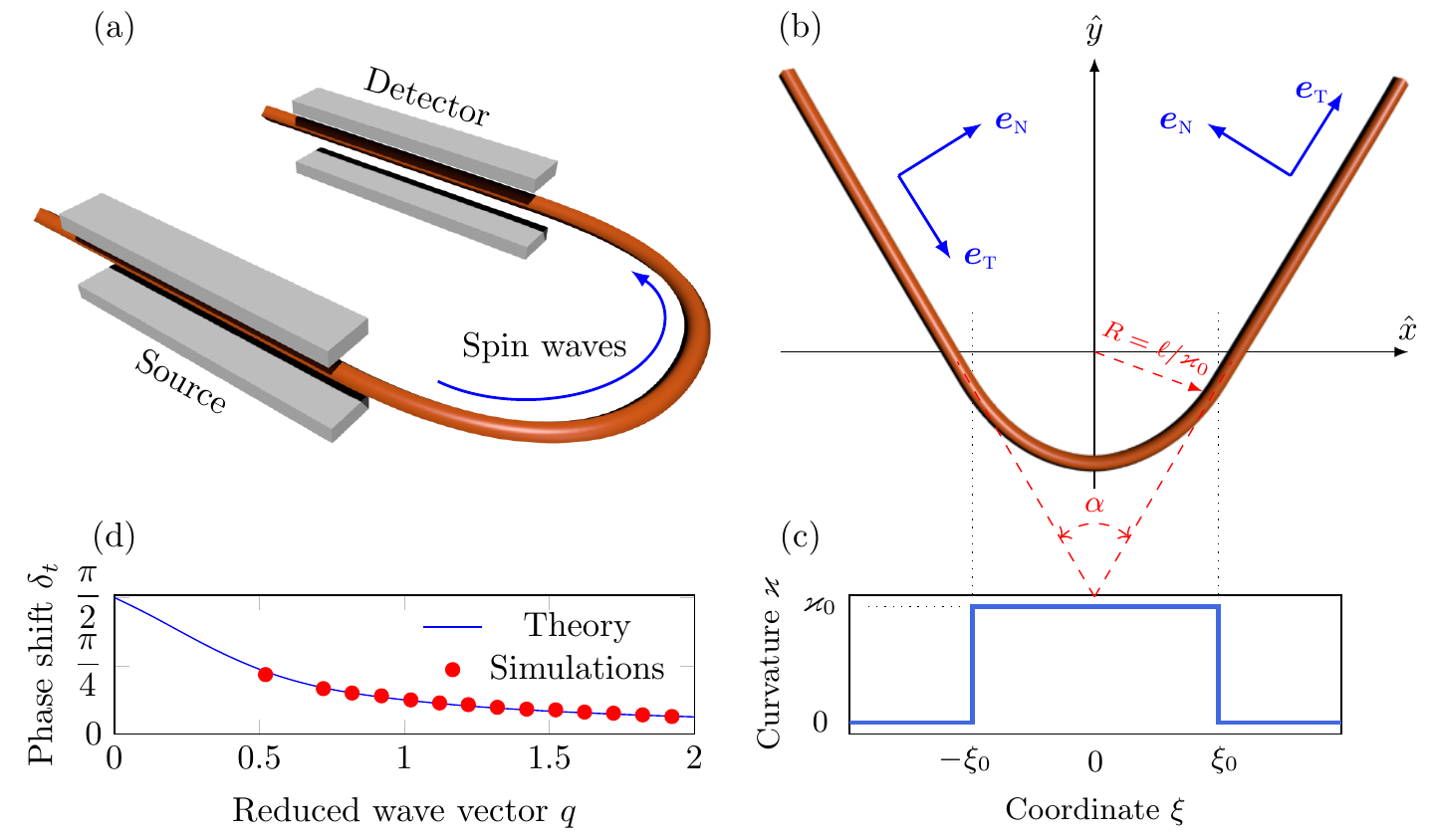}
	\end{center}
	\caption{
		\textbf{Curved flat ferromagnet wire:} (a) Schematic of the bent ferromagnet wire.
		The spin wave is generated by an alternating field inside the ``Source'' region and propagates through the bend to the detector. The phase of the transmitted wave is compared with the phase expected for the straight wire. 
		(b) Curve with a box-function curvature with $R$ being the bending radius and $\alpha$ being the angle of expansion. 
		(c) Box-function curvature profile. 
		(d) Phase shift of the transmitted spin wave according to equation \eqref{eq:carbox-eigenfunctions-scat} compared with spin-lattice simulations, see \eqref{eq:Born-1} for the case of small curvatures.}
	\label{fig:intro}
\end{figure*}

The purpose of the current study is to derive conditions when the bending of the wire results in localization of magnon states. Using a recently developed approach 
for the description of arbitrary curved wires \cite{Sheka15} we study the equilibrium magnetization state and spin waves on its background for flat curved nanowires. It is shown that for a smoothly curved flat nanowire the equilibrium state has an almost tangential magnetization distribution with small deviations which are proportional to the first derivative of curvature over the arc length. A truly local mode is shown to be always present for any infinitesimally weak localized curvature of the wire. This internal mode results in certain scattering features of the spin waves.

The paper is organized as follows. In Sec.~\ref{sec:model} the relevant model is formulated. We present a general description of the ground state and linear excitations in flat, curved ferromagnetic wires in Sec.~\ref{sec:general}. 
We illustrate the proposed approach using a wire with a box-car curvature, i.e. two straight segments connected by a circular arc in Sec.~\ref{sec:boxcar}. 
The bound states in the limiting case $\xi_0\to0$ of a sharp bend are considered in Sec.~\ref{sec:sharp}. Numerical simulations are described in Sec.~\ref{sec:num-sim} and concluding remarks are contained in Sec.~\ref{sec:conclusions}. Appendices contain the mathematical details of calculations, described in the main text.

\section{Model description}
\label{sec:model}

We base our study on the phenomenological Landau-Lifshitz equation
\begin{equation} \label{eq:LL}
\partial_t \vec m = \frac{\gamma_0}{M_s}\, \vec m\times\frac{\delta E}{\delta \vec m},
\end{equation}
which describes the precession dynamics of the classical magnetization unit vector
$\vec{m}$. Here $E$ is the total energy, $M_s$ is the saturation magnetization, and $\gamma_0 = g \mu_\textsc{b} / \hslash$ is the gyromagnetic ratio with $g$ being Land\'{e} $g$-factor, $\mu_\textsc{b}$ being Bohr magneton and $\hslash$ being Planck constant. The damping is neglected.

Let us consider a curved ferromagnetic wire of a shape given by an oriented plane curve $\vec{\gamma}(s)$ with a fixed cross-section of area $S$, parameterized by arc length $s$, see Fig.~\ref{fig:intro}(a).
It defines a basis in terms of tangential, normal and binormal directions $\vec{e}_{\textsc{t}}=\partial_s\vec{\gamma}$, $\vec{e}_{\textsc{n}}=\hat{\vec{z}}\times\vec{e}_{\textsc{t}}$, and $\vec{e}_{\textsc{b}}=\hat{\vec{z}}$, respectively. Here $\hat{\vec{z}}$ is a unit vector perpendicular to the plane containing the wire. In accordance with the Frenet--Serret formulae, the curvature $\kappa(s)$ of a planar wire is determined by the relation $\partial_s\vec{e}_\textsc{t}=\kappa\vec{e}_\textsc{n}$. Introducing the parameterization $\vec{e}_\textsc{t}=\cos\eta(s)\hat{\vec{x}}+\sin\eta(s)\hat{\vec{y}}$ one obtains $\kappa=\partial_s\eta$. Note that the curvature is positive (negative) for concave (convex) shapes, determined with respect to the $\hat{\vec{y}}$-axis.

The total magnetic energy of a sufficiently thin ferromagnetic wire of circular (or square) cross section is well--known \cite{Slastikov12}, and can be described by a reduced one-dimensional (1D) energy given by a sum of exchange and local anisotropy terms,
\begin{equation} \label{eq:Energy-1D}
E = S \int \mathrm{d}s \left[\mathcal{A}|\partial_s\vec{m}|^2 - K_\text{eff} \left(\vec{m}\cdot \vec{e}_{\textsc{t}}\right)^2\right],
\end{equation}
where $\mathcal{A}$ is the exchange constant and $K_\text{eff}=K+\pi M_s^2$ is the constant of the effective easy-tangential anisotropy, which incorporates the intrinsic crystalline anisotropy $K$ and the shape-induced magnetostatic contribution \cite{Slastikov12}, see also \cite{Gaididei17a}. The orientation of the easy-axis is determined by the coordinate-dependent tangential direction $\vec{e}_{\textsc{t}} = \vec{e}_{\textsc{t}}(s)$. This is a juncture where a sample geometry appears in the system: we assume that the anisotropy is strong enough to dictate the geometry-dependent magnetization configuration. By choosing the curvilinear reference frame adapted to the geometry we get rid of the coordinate dependence of the magnetic anisotropy term, hence it assumes its usual translation-invariant form. In the following we consider a magnet where such an easy-tangential anisotropy significantly exceeds all other interactions. It is convenient to express the magnetization in terms of the Frenet--Serret local reference frame, $\vec{m} = \sin\theta\cos\phi\,\vec{e}_{\textsc{t}}+\sin\theta\sin\phi\,\vec{e}_{\textsc{n}} + \cos\theta\,\vec{e}_{\textsc{b}}$, where the angular variables $\theta $ and $\phi $ depend on spatial and temporal coordinates. In the curvilinear reference frame the total energy \eqref{eq:Energy-1D} reads $E = E_0\int \mathscr{E}\mathrm{d}\xi$ with the energy density \cite{Sheka15}
\begin{equation} \label{eq:Energy-plane}
\mathscr{E} = {\theta'}^2 + \sin^2\theta \left(\phi'+\varkappa\right)^2 - \sin^2\theta\cos^2\phi.
\end{equation}
Here and below prime $'$ denotes the derivative with respect to the dimensionless coordinate $\xi=s/\ell$ with $\ell=\sqrt{\mathcal{A}/K_\text{eff}}$ being the characteristic magnetic length, the dimensionless curvature $\varkappa(\xi)=\ell \kappa(\xi)$, and $E_0 = S\sqrt{\mathcal{A}K_\text{eff}}$ being the characteristic energy of a domain wall in the rectilinear segment of the wire.

\section{General results: magnon eigenmodes for a weakly curved wire}

\label{sec:general}

Let us start our analysis with a static solution, which corresponds to the minimum of the energy functional \eqref{eq:Energy-plane}. The ground magnetization distribution for the planar curved wire is also planar: the magnetization lies in the wire plane (i.e. the osculating or TN-plane), hence the equilibrium state has the form 
\begin{subequations} \label{eq:Theta-Phi-statics}
	\begin{equation} \label{eq:gs}
	\Theta = \frac{\pi}{2},\quad \varPhi = \varPhi(\xi).
	\end{equation}
	The corresponding azimuthal magnetization angle $\varPhi$ is described by the equation of a nonlinear driven pendulum (see Appendix~\ref{sec:ground-state} for details):
	\begin{equation} \label{eq:LL-statics-phi}
	\varPhi '' - \sin\varPhi  \cos\varPhi  = -\varkappa'.
	\end{equation}
\end{subequations}
In the limiting case of a smoothly curved wire with localized curvature, the asymptotic solution of~\eqref{eq:LL-statics-phi} is valid:
\begin{equation} \label{eq:phi0lim}
\varPhi (\xi) = \varkappa'(\xi) + \mathcal{O} \left(\frac{|\varkappa'''|}{|\varkappa'|}\right),
\end{equation}
see Eqs.~\eqref{eq:smooth} for details. 


In order to analyze magnon modes we linearize the Landau--Lifshitz equations on the background of the static solution \eqref{eq:Theta-Phi-statics} by considering the small deviations $\psi=\theta-\Theta + i \left(\phi-\varPhi\right)$. Then the linearized Landau-Lifshitz equations can be presented in the form of a generalized Schr\"odinger equation, originally proposed for the description
of magnons on the magnetic vortex background in Ref.~\cite{Sheka04}, and used later for studying magnon modes (including local ones) over the precession soliton  \cite{Ivanov05b} and over the magnetic skyrmion \cite{Kravchuk18} in easy-axis magnets. Recently a similar approach was used for the description of the magnon spectrum in a curved helix wire \cite{Sheka15c}. For the case of the planar curved wire one gets
the following generalized Schr\"odinger equation, see Appendix \ref{sec:ground-state} for details:
\begin{subequations}
	\begin{equation} \label{eq:Schroedinger}
	-i \dot{\psi} = H\psi + W\psi^\star, \qquad H = -\partial_\xi^2 +1+V.
	\end{equation}
	Here the overdot indicates the derivative with respect to the dimensionless time $\tilde{t}=t\omega_0$ with $\omega_0=2\gamma_0K_\text{eff}/M_s$, the star operator means the complex conjugation, and the ``potentials'' have the following form:
	\begin{equation} \label{eq:V-n-W}
	\begin{split}
	V(\xi) &= -\frac{1}{2}\left[3\sin^2\varPhi  + \left(\varPhi ' + \varkappa\right)^2\right],\\
	W(\xi) &= \frac{1}{2}\left[\sin^2\varPhi  - \left(\varPhi ' + \varkappa\right)^2\right].
	\end{split}
	\end{equation}
\end{subequations}
One can see that the presence of these potentials is caused by the deviation of magnetization from the tangential direction; the functions $V$ and $W$ are well localized in the vicinity of the wire bend, i.~e. in the place where the curvature is present.

In order to analyze the generalized Schr\"{o}dinger equation~\eqref{eq:Schroedinger} we consider stationary states of the form
\begin{equation} \label{eq:Ansatz}
\psi(\xi,\tilde{t}) = u(\xi) e^{i{\varOmega\tilde{t}}} + v^\star(\xi) e^{-i{\varOmega\tilde{t}}}.
\end{equation}
Here $\varOmega = \omega/\omega_0 > 0$ is the normalized frequency. Thus we then finally obtain the following eigenvalue problem (EVP) for
the functions $u$ and $v$:
\begin{subequations} \label{eq:EVP}
	\begin{align} \label{eq:eq4u-k}
	&-u'' + \left(V + \varepsilon \right)u=-W v,\\
	\label{eq:eq4v-k} %
	&-v'' + \left(2 + V - \varepsilon \right)v = -W u.
	\end{align}
\end{subequations}
Here $\varepsilon$ describes the deviation of the eigenfrequency from the gap value, $\varOmega = 1 - \varepsilon$. Note that the function $v(\xi)$ has the asymptotic exponential behavior, $v(\xi)\propto e^{-\sqrt{2-\varepsilon}|\xi|}$ far from the bend region, hence one can always choose $v(\xi)$ as a slave function in the EVP \eqref{eq:EVP}. 

Let us discuss the EVP \eqref{eq:EVP} for $\varOmega>1$, i.~e. $\varepsilon<0$. In this case we get the  scattering problem for the complex-valued function $u(\xi)$ with the asymptotic scattering conditions
\begin{equation} \label{eq:scat-asymp}
u(\xi) = 
\begin{dcases*}
e^{iq\xi}+\mathcal{R}e^{-iq\xi}, & when $\xi\to-\infty$,\\
\mathcal{T}e^{iq\xi},            & when $\xi\to\infty$.
\end{dcases*}
\end{equation}
Here $\mathcal{R}=Re^{i\delta_r}$ and $\mathcal{T}=Te^{i\delta_t}$ are the complex amplitudes of reflected and transmitted magnons, respectively. The corresponding scattering phases are $\delta_r$ and $\delta_t$. 

Below in this section we analyze the EVP \eqref{eq:EVP} for the infinitesimally weak curvature, $|\varkappa|\ll1$. In this limiting case the effective potentials \eqref{eq:V-n-W} read
\begin{equation} \label{eq:V-n-W-as}
V(\xi) = W(\xi) = -\frac{\varkappa^2(\xi)}{2} + \mathcal{O}\left(\frac{|\varkappa'|}{|\varkappa|}\right).
\end{equation}
When the frequency $\varOmega > 1$, the solution for the straight wire reads
$u(\xi) \propto e^{iq \xi}$, with $v(\xi)=0$ and $\varOmega = 1+q^2$,
where $q=k\ell$ is the dimensionless wave number. Due to the curvature, the two equations for the magnon amplitudes, \eqref{eq:EVP}, become coupled.
Nevertheless, far from the bend region, the coupling potential $W$ is small. 
Neglecting all second order terms due to curvature in~\eqref{eq:EVP}, it gets the form of the usual Schr\"{o}dinger equation, see \eqref{eq:Scroedinger4u} for details. Then the scattering problem of magnons by the bending becomes equivalent to the scattering problem of a quantum particle by the potential $V(\xi)$ and the scattering boundary conditions \eqref{eq:scat-asymp}. In particular, in the case of a weak potential, $|V|\ll1$, the Born approximation is valid, 
\begin{equation} \label{eq:Born}
\tan\delta{_t} = -\frac{1}{2\sqrt{\varepsilon}}\int_{-\infty}^{\infty} V(\xi)\mathrm{d}\xi,
\end{equation} 
which results in the following phase shift
\begin{equation} \label{eq:Born-1}
\delta{_t}(q) = \arctan \frac{\varDelta}{q}, \qquad \varDelta = \frac{1}{4} \int_{-\infty}^{\infty} \varkappa^2(\xi)\mathrm{d}\xi.
\end{equation} 
Note that $\delta{_t} \approx \pi/2 - q/\varDelta$ for $q \ll1 $ and $\delta{_t} \approx \varDelta/q$ for $q \gg 1$.

Now let us consider the local modes (bound states) inside the gap with the frequency $\varOmega \in (0,1)$, i.~e. $\varepsilon \in (0,1)$. In this case both functions $u$ and $v$ become localized. One can always choose $u(\xi)$ as a real-valued function. The function $u$ has the asymptotic behavior of the form $u(\xi)\propto e^{-\sqrt{\varepsilon}|\xi|}$.

Under the same assumptions, the EVP is equivalent to the quantum-mechanical problem about the energy eigenvalue of a quantum particle in a shallow well. It is known that the Hamiltonian $H_\lambda \equiv -\partial_\xi^2 +\lambda V(\xi)$ always has a bound state for all small positive $\lambda$ if the potential $V(\xi)$ is negative on average, i.~e. $\int_{-\infty}^\infty V(\xi) \mathrm{d}\xi<0$. The corresponding eigenvalue $\varepsilon$ is determined as follows \cite{Simon76,LandauIII}   
\begin{equation} \label{eq:energy-bound}
\sqrt{\varepsilon} = \frac{\lambda}{2} \int_{-\infty}^{\infty} V(\xi) \mathrm{d}\xi + \mathcal{O} \left(\lambda^2\right).
\end{equation}
The corresponding frequency of the local mode and the eigenfunction are 
\begin{subequations} \label{eq:omega-local}
	\begin{equation} \label{eq:energy-local}
	\begin{split}
	\varOmega^{\text{loc}} &\approx 1-\varDelta^2,\\
	u^{\text{loc}}(\xi) &\approx 
	\begin{dcases*}
	u_0, & inside the well,\\
	u_0\exp\left(-\varDelta|\xi|\right), & outside the well,
	\end{dcases*}
	\end{split}
	\end{equation}
	where $u_0$ is a normalized constant and $\varDelta$ is determined by \eqref{eq:Born-1}. In physical units 
	\begin{equation} \label{eq:omega-loc}
	\omega^{\text{loc}} = \omega_0 \left\{1 - \frac{1}{16} \left[
	\int_{-\infty}^{\infty} \varkappa^2(\xi)\mathrm{d}\xi\right]^2\right\}.
	\end{equation}
\end{subequations}

According to the quantum scattering theory, the presence of bound states influences the scattering data. This influence is given by the Levinson theorem~\cite{Swan63}: the total phase shift $\delta_t(0)-\delta_t(\infty)$ is related to the number of bound states (i.e. local modes) $N^{\text{loc}}$ and half--bound states (i.e. half-local modes) $N^{\text{h.loc}}$ \cite{Ma06a}. In the 1D case the total phase shift reads \cite{Dong00b}
\begin{equation} \label{eq:Levinson}
\!\!\!
\frac{\delta{_t}(0)-\delta{_t}(\infty)}{\pi} = \!
\!
\begin{dcases*}
N_e^{\text{loc}} \!\!+ \frac12 N_e^{\text{h.loc}} \!\! -\frac12,\!\! & even parity,\!\!\\
N_o^{\text{loc}} + \frac12 N_o^{\text{h.loc}}, 			& odd parity.\!\!
\end{dcases*}
\end{equation}
We will use the Levinson theorem in the form \eqref{eq:Levinson} in order to check the number of local modes.

\section{Magnon eigenmodes for a wire with a box curvature}
\label{sec:boxcar}

\begin{figure}
	\begin{center}
		\includegraphics[width=\columnwidth]{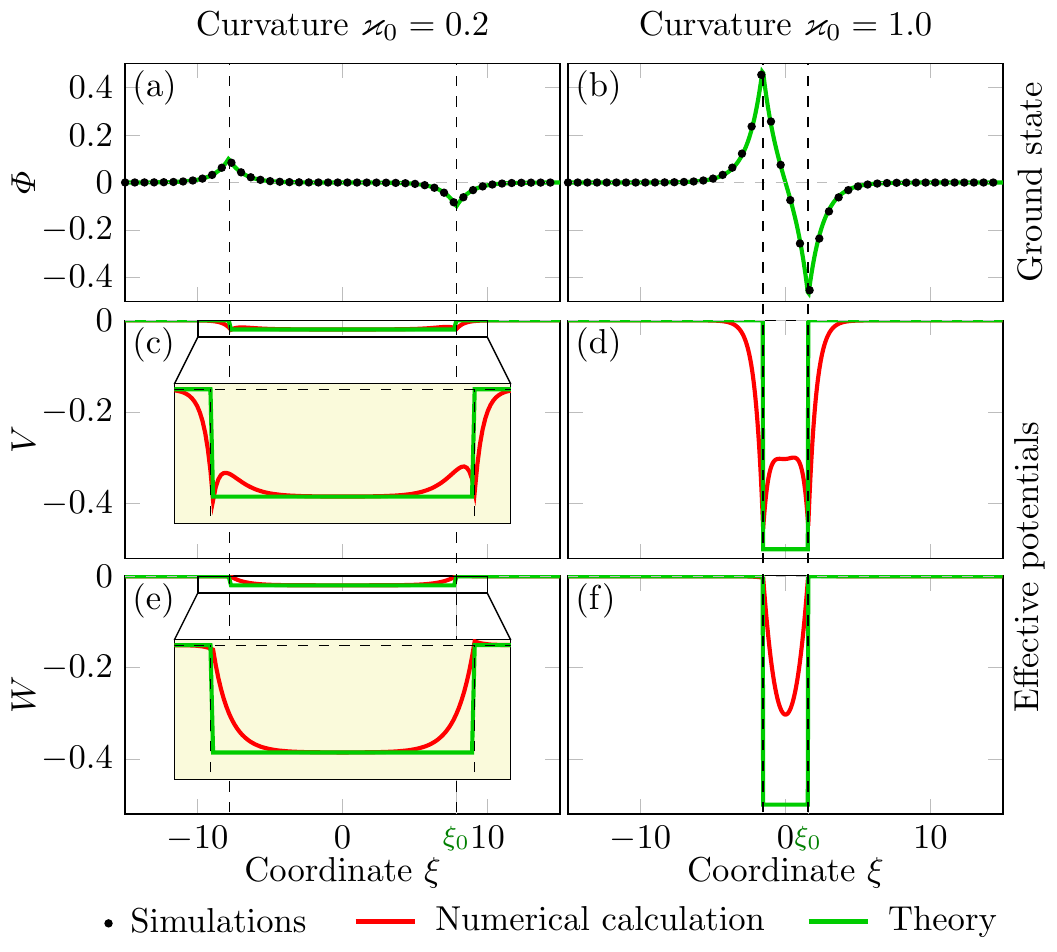}
	\end{center}
	\caption{\textbf{Magnetization distribution for a wire with a box curvature:}(a), (b) Equilibrium magnetization states for the wire with the box curvature for different curvature values $\varkappa_0$. 
		Symbols correspond to simulations and solid lines to analytics~\eqref{eq:boxcar-gs-linear}. 
		(c)--(f) Effective potentials $V$ and $W$ for different curvatures (left and right columns). 
		Red curves correspond to the expressions~\eqref{eq:V-n-W} according to the magnetization equilibrium state~\eqref{eq:boxcar-gs}. 
		Green curves show the box approximation of the potentials for the analytics~\eqref{eq:match-varepsilon}.}
	\label{fig:ground-state}
\end{figure}

Let us consider a simple model of the curve, which has a constant curvature $\varkappa _0$ in the interval $s\in[-\xi_0,\xi_0]$ and vanishes outside (box--function curvature):
\begin{equation} \label{eq:kappa4arc}
\varkappa (s) = \varkappa _0 h(\xi+\xi_0) - \varkappa _0 h(\xi-\xi_0),
\end{equation}
where $h(\xi)$ is the Heaviside step function, $\varkappa_0={\ell}/R$ is the curvature of the arc bending with $R$ being the curvature radius, and $\xi_0 = (\pi-\alpha)R/(2{\ell})$ with $\alpha$ being the expansion angle, the angle between the non-bending part of the curve, see Figs.~\ref{fig:intro}(b),~(d). If we consider the non-intersecting curve, then $\alpha\in (0,\pi)$.

In the limiting case of weak curvature, $|\varkappa_0| \ll 1$, the ground state has a form of two hooks in the points of curvature step with exponential tails:
\begin{equation} \label{eq:boxcar-gs-linear}
\varPhi (\xi) =
\begin{dcases*}
-\varkappa _0 e^{-\xi_0} \sinh\xi, & when $|\xi|\leq \xi_0$,\\
-\text{sgn}(\xi) \varkappa _0 e^{-|\xi|} \sinh\xi_0, &when $|\xi|> \xi_0$,
\end{dcases*}
\end{equation}
see Figs.~\ref{fig:ground-state}(a),~(b). Then the effective potentials~\eqref{eq:V-n-W-as} are
\begin{equation} \label{eq:V-n-W-carbox}
V(\xi) \approx W(\xi) \approx 
\begin{dcases*}
-\frac{\varkappa_0^2}{2}, &when $|\xi| < \xi_0$,\\
0, &when $|\xi| > \xi_0$,
\end{dcases*}
\end{equation}
see Figs.~\ref{fig:ground-state}(c)--(f) for comparison with the exact form~\eqref{eq:V-n-W}.

\begin{figure*}
	\begin{center}
		\includegraphics[width=\linewidth]{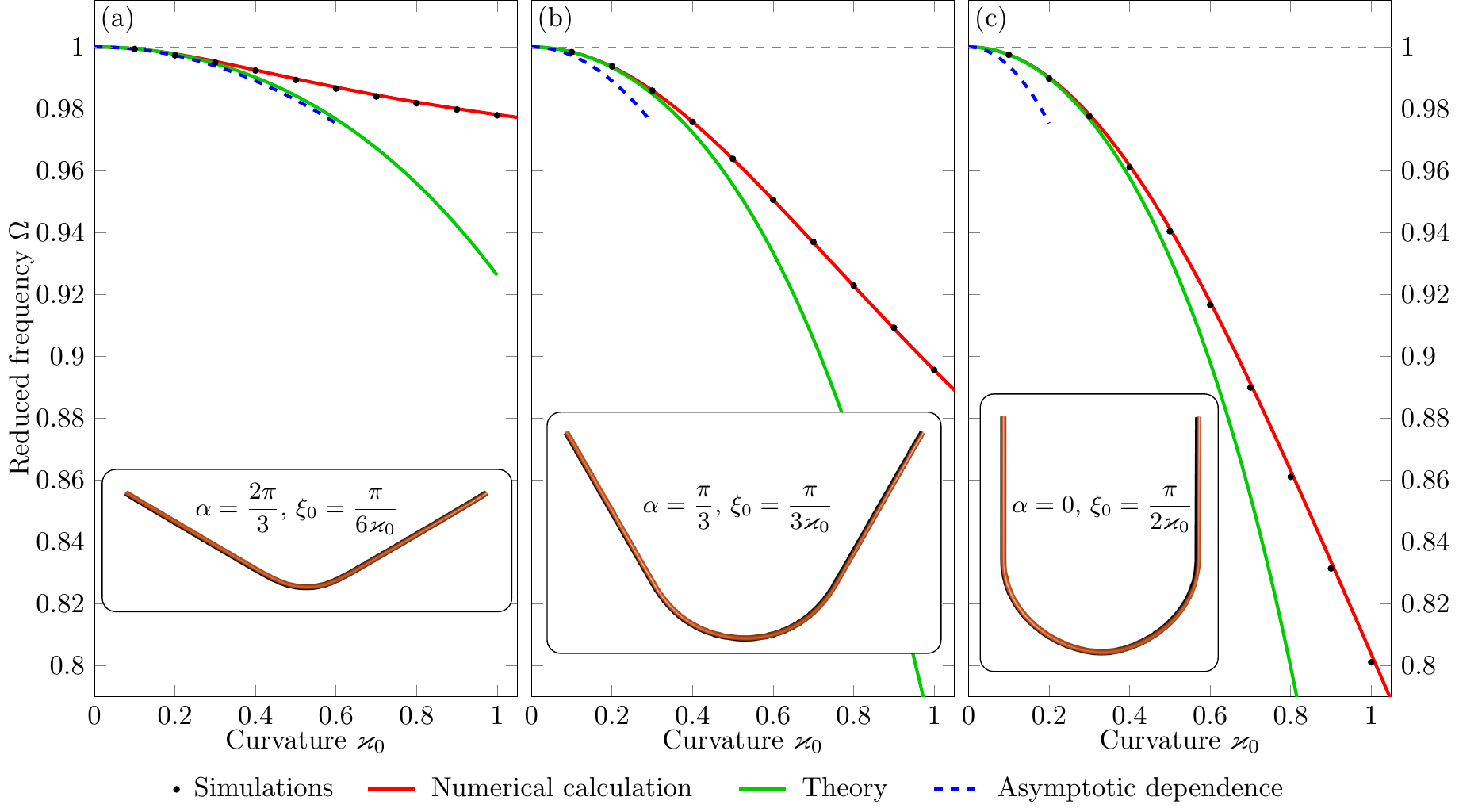}
	\end{center}
	\caption{
		\textbf{Local mode for a wire with a box curvature:} Dependencies of local mode's frequency $\varOmega$ on the curvature $\varkappa_0$ for magnetic nanowires with different arc lengths, namely, (a) for $\xi_0=\pi/(6\varkappa_0)$, (b) for $\xi_0=\pi/(3\varkappa_0)$ and (c) for $\xi_0=\pi/(2\varkappa_0)$. 
		Symbols correspond to the simulation results, the red curves represent the numerical solutions of the eigenvalue problem \eqref{eq:EVP} for different curvatures, the green curves show the numerical solution of the dispersion equation~\eqref{eq:match-varepsilon}, and the blue curves correspond to the asymptotic dependences, according to the limiting cases \eqref{eq:energy-carbox-narrow} and \eqref{eq:energy-carbox-half-circle}.}
	\label{fig:local-mode}
\end{figure*}

Let us consider the EVP \eqref{eq:EVP} with box potentials \eqref{eq:V-n-W-carbox}. 
For the shallow well ($\varkappa_0\ll 1$), one can find
\begin{equation} \label{eq:energy-carbox-narrow}
\varepsilon\approx \frac{\varkappa_0^4 \xi_0^2}{4}, \quad \text{when $\varkappa_0\ll1$},
\end{equation}
which is in agreement with \eqref{eq:energy-local}. In the particular case of the wide well, namely for the limiting case of a half--circumference arc ($\alpha=0$), one gets
\begin{equation} \label{eq:energy-carbox-half-circle}
\varepsilon\approx \frac{\pi^2 \varkappa_0^2}{16}, \quad \text{when $\varkappa_0\ll1$ and $\xi_0 =\frac{\pi}{2\varkappa_0}\gg1$}.
\end{equation} 

Eigenfrequencies of local modes for different geometrical parameters are shown in Fig.~\ref{fig:local-mode}.
While the asymptotics work well only for narrow wells, the exact analytical solution for a box potential~\eqref{eq:V-n-W-carbox} shows a good agreement with simulations for $\varkappa_0 \lesssim 0.4$. 
The numerical solution for the initial EVP~\eqref{eq:EVP} perfectly agrees with the simulations in the whole range of the calculated data.

The scattering problem corresponds to the EVP~\eqref{eq:EVP} for $\varOmega > 1$. We are mainly interested in the forward scattering of the transmitted wave. The scattering phase $\delta_t = \arg \mathscr{T}$, calculated using \eqref{eq:carbox-scat}, is plotted in Fig.~\ref{fig:intro}(d).

\section{Bound states on a sharp bend}
\label{sec:sharp}

\begin{figure}
	\includegraphics[width=\columnwidth]{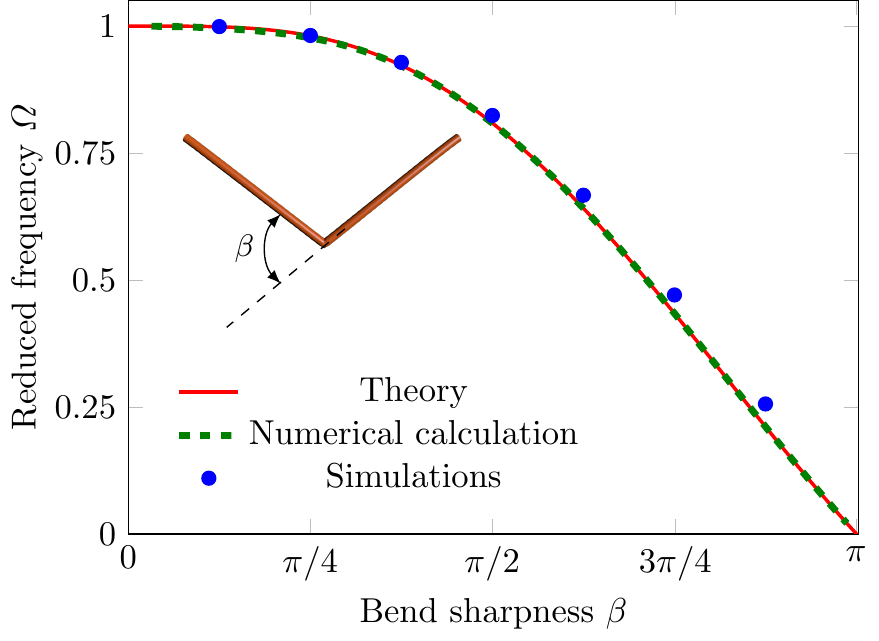}
	\caption{
		\textbf{Local mode on a sharp bend:} 
		Eigenfrequency of the mode localized on a sharp bend depending on the bend angle $\beta=\pi-\alpha$. Solid line is determined by (\ref{eq:Omega}), dashed line shows the numerical solution of the EVP \eqref{eq:EVP} and markers correspond to spin-lattice simulations. Inset shows geometry of the wire.}
	\label{fig:Omega-delta}
\end{figure}

Here we consider a limiting case of a sharp bend when $\xi_0\to0$ with the fixed opening angle $\alpha$, see Figs.~\ref{fig:intro}(b),~(c). In this case, taking into account the relation $\varkappa_0=(\pi-\alpha)/(2\xi_0)$ one obtains
\begin{equation} \label{eq:delta}
\varkappa = \beta\delta(\xi),
\end{equation}
where $\beta=\pi-\alpha$ and $\delta(\xi)$ is the Dirac delta-function. The conditions of a smoothly curved wire \eqref{eq:smooth-conditions} are not valid here. Therefore we start with the initial equation \eqref{eq:LL-statics-phi}, which for the sharp bend reads
\begin{equation}\label{eq:Phi-delta}
\varPhi''-\sin\varPhi\cos\varPhi=-\beta\delta'(\xi).
\end{equation}
For the boundary conditions $\varPhi(\pm\infty)=0$ Eq.~\eqref{eq:Phi-delta} has a solution
\begin{equation}\label{eq:Phi-solution}
\varPhi(\xi)=-2\,\mathrm{sgn}(\xi)\arctan e^{-|\xi|+\xi_c},\quad \xi_c=\ln\tan\frac{\beta}{4},
\end{equation}
for details see Appendix~\ref{app:sharp}. Substituting \eqref{eq:Phi-solution} into \eqref{eq:V-n-W} one obtains the potentials in the form
\begin{equation}\label{eq:V-W-delta}
V=-\frac{2}{\cosh^2(-|\xi|+\xi_c)},\qquad W=0.
\end{equation}
Finally, the EVP is presented by two independent homogeneous equations \eqref{eq:EVP-delta}. Let us first consider the localized states with $\varOmega<1$. In this case Eqs.~\eqref{eq:EVP-delta} have the bound solution
\begin{subequations}\label{eq:u-v}
	\begin{align}
	\label{eq:uu}
	&u=C\,e^{-\sqrt{\varepsilon}|\xi|}\left[\sqrt{\varepsilon}+\tanh(|\xi|-\xi_c)\right],\\
	\label{eq:vv} &v=0,
	\end{align}
\end{subequations}
and $\sqrt{\varepsilon}=f(\beta)$, see \eqref{eq:f}. The corresponding eigenfrequency $\varOmega=1-\varepsilon$ reads 
\begin{equation}\label{eq:Omega}
\varOmega=\frac{1}{2}\cos\frac{\beta}{2}\left[\cos\frac{\beta}{2}+\sqrt{4-3\cos^2\frac{\beta}{2}}\right].
\end{equation}
For details see Appendix~\ref{app:sharp}. The dependence $\Omega(\beta)$ is shown in Fig.~\ref{fig:Omega-delta}

Note a perfect agreement of the analytical (solid line) and numerical (dashed) results. The latter were obtained as a numerical solution of the EVP \eqref{eq:EVP}, where the $\delta$-function on the discrete lattice was modeled as
\begin{equation}\label{eq:delta-appr}
\delta(\xi_i)=
\begin{dcases*}
\frac{1}{\Delta\xi},& when $\xi_i=0$,\\
0,	&when $\xi_i\ne 0$,
\end{dcases*}
\end{equation}
where $\Delta\xi=10^{-2}$ is the step of the discretization. The equilibrium value of the function $\varPhi(\xi_i)$ was found by means of a numerical minimization of the energy \eqref{eq:Energy-plane} with $\theta=\pi/2$. 

For the case $\varOmega>1$ one obtains a scattering problem. The solution of   Eq.~\eqref{eq:u-delta} can be presented in the form of incident, transmitted and reflected waves, see \eqref{eq:u-scater}. The corresponding coefficients of transmission and reflection read
\begin{equation}\label{eq:T}
T^2=\frac{(\varOmega-1)\varOmega^2}{\cos^2\frac{\beta}{2}\sin^4\frac{\beta}{2}+(\varOmega-1)\varOmega^2}
\end{equation}
and $R^2=1-T^2$, respectively. For details see Appendix~\ref{app:sharp}.
Note that the full transition takes place for the angles $\beta=0$ (rectilinear wire) and $\beta=\pi$ (completely bent wire), while the minimum transition corresponds to the angle $\beta_c=2\arccos(1/\sqrt{3})$.

The total phase shift between the transmitted and incident waves is $\delta_{\text{tot}}(q)=\delta_t(q)+2\arctan(1/q)$, where $q=\sqrt{\varOmega-1}$ is the dimensionless wave-vector and $\delta_t(q)$ is determined in \eqref{eq:delta_t}. Since $\delta{_t}(0)-\delta{_t}(\infty)=\pi/2$, the Levinson theorem \eqref{eq:Levinson} results in a single bound state.

\section{Spin-lattice simulations of the magnon dynamics}
\label{sec:num-sim}

Our predictions about the localized mode statics and dynamics are verified by performing 3D spin-lattice simulations using the in-house developed program \textsf{SLaSi}~\cite{SLaSi}. We model the ferromagnetic nanowire with the local box curvature [see Fig.~\ref{fig:intro}(b)] as a classical chain of magnetic moments $\vec{m}_i$, with $i=\overline{1,N}$. 
Neighboring moments interact according to the anisotropic Heisenberg model Hamiltonian as follows:
\begin{equation} \label{eq:hamiltonian}
\!\!\mathscr{H} = - a\ell^2 \!\!\sum_{i=1}^{N-1}\!\! \vec{m}_i\vec{m}_{i+ 1}  -  a^3 \! \sum_{i=1}^N \!\!\left[(\vec{m}_i \vec{e}_{\textsc{t}i})^2 \! + \vec{h}_i \vec{m}_i \right]\!\!.
\end{equation}
Here $a$ is a lattice constant and $\vec{h}_i \equiv \vec{h}(\vec{r}, t)$ is a magnetic field being a function of coordinate and time, measured in units of the coercive field $H_c=K_{\text{eff}}/M_s$.
The dynamics of this system is described by a set of $N$ vector Landau--Lifshitz--Gilbert ordinary differential equations, see Ref.~\cite{Pylypovskyi14} for a general description of the \textsf{SLaSi} simulator and Ref.~\cite{Sheka15c} for the details of the simulations of curved wires. 

To study the equilibrium magnetization states we consider a chain of $N=3000$ sites with an exchange length $\ell=15 a$. 
The box curvature part of the wire is placed in the center of the chain with the arc length being $\xi_0 = (\pi - \alpha)R/(2\ell)$, where $\alpha$ is the angle between no-bending parts of the wire, which take the following values $(0,\pi/6,\pi/3,\pi/2)$.
The curvature of the central part of the wire is varied in a wide range of parameters $\varkappa_0 \in (0,1.0)$, with a step $\Delta \varkappa_0 = 0.1$. 
In order to verify our theoretical prediction, we perform our simulation starting from the tangential magnetic distribution.
We simulate numerically the set of $N$ discrete Landau--Lifshitz--Gilbert equations in the overdamped regime with the Gilbert damping $\eta=0.5$ during the long-time interval $\Delta t \gg (\eta \, \omega_0)^{-1}$. The final state with the lowest energy is considered to be the equilibrium magnetization state. Simulation data are shown in the Figs.~\ref{fig:ground-state}(a),~(b) by filled symbols together with theoretical results (solid lines) for small and high curvature. 

Figures~\ref{fig:ground-state}(c)--(f) show a comparison of the exact potentials~\eqref{eq:V-n-W} with their analytical approximation.The direct numerical calculation of the eigenmodes of~\eqref{eq:EVP} precisely agrees with results of spin-lattice simulations in a wide range of curvatures, see Fig.~\ref{fig:local-mode}(g) and Fig.~\ref{fig:Omega-delta}. Frequencies of local modes $\varOmega$ are calculated in the following way. After obtaining the equilibrium magnetic state in the system, a pulse of a homogeneous magnetic field along $\hat{y}$ axis is applied to the system. Due to its inhomogeneity in the local curvilinear frame of reference, this pulse generates a wide spectrum of spin waves as well as local modes, which is situated inside the gap of the spin-wave spectrum. 

The numerical verification of the Levinson theorem \eqref{eq:Levinson} is a way to check the number of bound states inside the bend. We perform simulations of the spin wave scattering on a spin chain of 14\,000 magnetic moments with a localized curvature $\varkappa_0 = 0.5$ and $\xi_0 = \pi$, see Fig.~\ref{fig:intro}(c). A spatially homogeneous, time alternating magnetic field is applied between the 1000-th and the 6000-th chain sites to generate a spin wave propagating through the bend. To determine the phase shift $\delta_t$ due to the bend, we compare the spin wave propagation in two different wires with the same material parameters: one is a curved wire and another one is a straight wire of the same length. Spatial magnetization distribution of a given time moment is analyzed in a detector window [see schematics in Fig.~\ref{fig:intro}(a)]. The detector is placed on the same arc length from the source in both wires. The resulting phase shift is plotted in Fig.~\ref{fig:intro}(d). It is clearly seen that there exists one localized mode in the bend.

\section{Conclusions}
\label{sec:conclusions}

In conclusion, the general analysis of the ground state magnetization and spin wave spectrum is performed in the case of a very small curvature of the wire. The equilibrium magnetization distribution deviates from the tangential wire direction only in the region of the wire bend, the corresponding deviation angle is determined only by the spatial derivative of the curvature, resulting in an effective potential well for spin waves. We predict the existence of a curvature induced truly local mode, localized on a bend, with an eigenfrequency below the ferromagnetic resonance. The interaction of the local mode with spin waves, propagating through the bend, results in a shift of the wave phase according to the Levinson theorem. In the limiting case of a sharp bend ($\delta$-function curvature) the analytically exact expressions for eigenfrequencies of the bound states, as well as for transmission and reflection coefficients of the scattering process are found.

\section{Acknowledgments}

D.~D.~S. and O.~V.~P. thank the University of Bayreuth and Helmholtz-Zentrum Dresden-Rossendorf e.~V. (HZDR), where part of this work was performed, for their kind hospitality and acknowledge the support from the Alexander von Humboldt Foundation (Research Group Linkage Programme). V.~P.~K. acknowledges the support from the Alexander von Humboldt Foundation. All simulations for this work were performed using the computer clusters of the Bayreuth University~\cite{Bayreuth_cluster}, Taras Shevchenko National University of Kyiv~\cite{unicc}, Bogolyubov Institute for Theoretical Physics of the National Academy of Sciences of Ukraine~\cite{bitpcluster} and HZDR~\cite{hypnos}. A.~S. was supported by the US Department of Energy. Authors thank D.~Makarov (HZDR) for fruitful discussions.

\appendix

\section{Ground state and spin wave spectrum}
\label{sec:ground-state}

The equilibrium magnetization states satisfy the energy minimum:
\begin{equation} \label{eq:LL-statics}
\begin{split}
&\Theta'' - \sin\Theta\cos\Theta \left[\left(\varPhi'+\varkappa\right)^2-\cos^2\varPhi\right]=0,\\
&\varPhi'' +2\Theta'\left(\varPhi' + \varkappa \right)\cot\Theta - \sin\varPhi \cos\varPhi +\varkappa'=0.
\end{split}
\end{equation}
For a plane curve, $\Theta = \pi/2$ and the second equation of~\eqref{eq:LL-statics} takes the form \eqref{eq:LL-statics-phi}.

The limiting case of a smoothly curved wire with a localized curvature corresponds to the following conditions:
\begin{subequations} \label{eq:smooth}
	\begin{equation} \label{eq:smooth-conditions}
	\begin{split}
	&|\varkappa|,|\varkappa'|\ll1, \qquad |\varkappa^{(n+1)}|< |\varkappa^{(n)}|, \; n\in\mathbb{N},\\
	&\varkappa(\pm\infty) = \varkappa'(\pm\infty)=0.
	\end{split}
	\end{equation}
	In this case the linearization of \eqref{eq:LL-statics-phi} with respect to $\varPhi $ gives the solution
	\begin{equation} \label{eq:smooth-equation}
	\varPhi (\xi) = \frac12\int_{-\infty}^{\infty} \varkappa'(\xi+\eta) e^{-|\eta|} \mathrm{d}\eta = \sum_{n=0}^\infty \varkappa^{(2n+1)} (\xi).
	\end{equation}
	In the main approach the asymptotic value~\eqref{eq:phi0lim} is valid under the condition $|\varkappa'''|\ll |\varkappa'|$.
\end{subequations}

In order to analyze magnon modes we linearize the Landau--Lifshitz Eq.~\eqref{eq:LL} on the background of the ground state \eqref{eq:gs} by considering the small deviations $\vartheta=\theta-\pi/2$ and $\varphi = \phi-\varPhi $. The energy of magnons in the harmonic approximation reads
\begin{equation} \label{eq:energy-magnons}
\mathscr{E}^{\text{mag}} = \vartheta'^2 + \varphi'^2 + V_1 \vartheta^2 + V_2 \varphi^2,
\end{equation}
with the ``potentials''
\begin{equation} \label{eq:potentials}
\begin{split}
V_1(\xi) &= \cos^2\varPhi (\xi) - \left[\varPhi '(\xi)+\varkappa(\xi)\right]^2, \\
V_2(\xi) &= \cos2\varPhi (\xi).
\end{split}
\end{equation}
The linearized Landau-Lifshitz equations can be presented in the following form
\begin{equation} \label{eq:LL-linearized}
-\vartheta'' + V_1 \vartheta = \dot \varphi,  \qquad -\varphi''   + V_2 \varphi   =- \dot \vartheta,
\end{equation}
where an overdot indicates the derivative with respect to the reduced time $\tilde{t} = \omega_0 t$. One can combine the set of equations \eqref{eq:LL-linearized} in a single equation for the complex-valued function $\psi = \vartheta+i\varphi$, which has the form of the generalized Schr\"odinger equation \eqref{eq:Schroedinger}
\begin{equation} \label{eq:Schroedinger-1}
\tag{\ref{eq:Schroedinger}$'$}
-i \dot{\psi} = H\psi + W\psi^\star, \qquad H = -\partial_\xi^2 +U.
\end{equation}
The corresponding ``potentials'' in Eq.~\eqref{eq:Schroedinger-1} read
\begin{equation} \label{eq:U-and-W}
U = \frac{V_1+V_2}{2}, \qquad W = \frac{V_1-V_2}{2}.
\end{equation}
Since we are interested in the stationary states solution of the form
\eqref{eq:Ansatz}, we get an EVP \eqref{eq:EVP}. The boundary conditions essentially depend on the type of the problem. For the scattering problem ($\varOmega>1$) one gets traveling waves together with localized modes with the following asymptotic behavior:
\begin{equation}
\psi \asymp
\begin{dcases*}
e^{i \left(q\xi + \varOmega \tilde{t}\right)} +\mathcal{R} e^{-i\left(q\xi - \varOmega \tilde{t}\right)}\!\!\\ 
+ c_1 e^{\sqrt{2-\varepsilon}\xi} e^{-i\varOmega \tilde{t}},\!\!& when $\xi\to-\infty$,\\
\mathcal{T}e^{i \left(q\xi + \varOmega \tilde{t}\right)}\\
+ c_2 e^{-\sqrt{2-\varepsilon}\xi} e^{-i\varOmega \tilde{t}},\!\! & when $\xi\to+\infty$.
\end{dcases*}
\end{equation}
In the case of the local mode ($\varOmega<1$) the asymptotic behavior corresponds to exponentially localized oscillations:
\begin{equation}
\psi \asymp
\begin{dcases*}
c_1 e^{\sqrt{\varepsilon}\xi} e^{i\varOmega \tilde{t}}
+ c_2 e^{\sqrt{2-\varepsilon}\xi} e^{-i\varOmega \tilde{t}},\!\!&	when $\xi \to -\infty$,\\
c_3 e^{-\sqrt{\varepsilon}\xi} e^{i\varOmega \tilde{t}}
+ c_4 e^{-\sqrt{2-\varepsilon}\xi} e^{-i\varOmega \tilde{t}},\!\!	& when $\xi \to +\infty$.
\end{dcases*}
\end{equation}
Here $c_i$, $i=\overline{1,4}$ are some real amplitudes.

For further analysis it is convenient to shift the potential $U$ by its asymptotic value, $U = 1+V$. Then both potentials $V(\xi)$ and $W(\xi)$ are localized. Suppose that we are working under smooth conditions \eqref{eq:smooth}, then $|V|, |W|\ll1$. One can conclude from Eq.~\eqref{eq:eq4v-k} that $v\approx -\frac12 Wu$, therefore the function $v$ becomes the slave variable in Eqs.~\eqref{eq:EVP}. Using the explicit asymptotic form \eqref{eq:phi0lim} for $\varPhi(\xi) =\varkappa'(\xi)$, one gets Eq.~\eqref{eq:V-n-W-as}. Now, by neglecting the function $v\propto \varkappa^2 u$, we get from \eqref{eq:EVP} the usual Schr\"odinger equation of the following form
\begin{equation} \label{eq:Scroedinger4u}
-u'' +V(\xi) u = \varepsilon u, \qquad V(\xi) = -\frac{\varkappa^2(\xi)}{2}.
\end{equation} 

\section{Wire with a box curvature}

The equilibrium magnetization state, according to \eqref{eq:LL-statics-phi}, is described by the following ODE
\begin{equation} \label{eq:boxcar-ode}
\begin{split}
&\varPhi ''-\sin\varPhi \cos\varPhi  = - \varkappa'(\xi),\\
&\varkappa'(\xi) = \varkappa_0\delta (\xi+\xi_0) - \varkappa _0 \delta (\xi-\xi_0), \\
&\varkappa_0 = \kappa_0 {\ell}, \qquad \xi_0 = \frac{s_0}{{\ell}}.
\end{split}
\end{equation}
The solution of this model, which satisfies the boundary conditions $\varPhi  (\pm \infty)=0$, has the form
\begin{equation} \label{eq:boxcar-gs}
\!\!\!\!\varPhi  (\xi) =\!
\begin{dcases*}
\dfrac{\pi}{2} - \mathrm{am}\left(\frac{\xi}{k} + \mathrm{K}(k),k\right)\!\!,& when $|\xi|\leq \xi_0$,\\
-2\text{sgn}(\xi) \arctan e^{-|\xi|-\xi_1}\!\!, &when $|\xi|> \xi_0$.
\end{dcases*}
\end{equation}
Here $\mathrm{am}(x,k)$ is the Jacobi's amplitude and $\mathrm{K}(k)$ is the complete elliptic integral of the first kind \cite{NIST10}. Parameters $\xi_1$ and $k$ are determined by the matching conditions:
\begin{equation} \label{eq:matching}
\left[\varPhi\right]_{\xi_0}=0, \qquad \left[\varPhi'\right]_{\xi_0} = \varkappa_0,
\end{equation}
where $\left[\cdots\right]_{\xi_0} \equiv ( \cdots ) \bigr|_{\xi_0+0} - ( \cdots )\bigr|_{\xi_0-0}$. The modulus $k$ of the elliptic functions and the parameter $\xi_1$ can be calculated from the following conditions
\begin{equation} \label{eq:modulus-k}
\begin{split}
&\mathrm{dn}\left(\frac{\xi_0}{k},k\right) = \frac{2\varkappa_0 k k'}{\varkappa_0^2 k^2-k'^2},\\
&\sech \left(\xi_0+\xi_1\right) = -k' \mathrm{sd}\left(\frac{\xi_0}{k},k\right),\\
\end{split}
\end{equation}
where $k' = \sqrt{1-k^2}$ is the complementary modulus.

In particular, if the reduced curvature $\varkappa_0\ll1$, then one can use the expansion
\begin{equation} \label{eq:modulus-k-series}
k = 1-\frac{\varkappa_0^2}{2}e^{-2\xi_0} + \mathcal{O}\left(\varkappa_0^2\right).
\end{equation}
In this case the static solution \eqref{eq:boxcar-gs} takes the simpler form \eqref{eq:boxcar-gs-linear}.
The maximum value of $|\varPhi|$ is 
\begin{equation} \label{eq:max-Phi}
\max |\varPhi| \approx 
\begin{dcases*}
\varkappa_0 \xi_0, & when $\xi_0\ll1$,\\
\varkappa_0/2, &when $\xi_0\gg1$.
\end{dcases*}
\end{equation}

Let us consider the EVP \eqref{eq:EVP} with box potentials \eqref{eq:V-n-W-carbox}. We start the problem of magnon states with the local modes problem. In this case we look for the eigenfunctions in the following form:
\begin{equation} \label{eq:carbox-eigenfunctions}
\begin{split}
\!\!u(\xi)&=
\begin{dcases*}
u_1\cosh k_1\xi + u_2\cos q_1\xi, &	when $|\xi|<\xi_0$,\\
u_3 \exp\left(-k_2\left[|\xi|-\xi_0\right]\right), &	when $|\xi|>\xi_0$,
\end{dcases*}
\\
\!\!v(\xi)&=
\begin{dcases*}
v_1\cosh k_1\xi + v_2\cos q_1\xi, &	when $|\xi|<\xi_0$,\\
v_3 \exp\left(-k_3\left[|\xi|-\xi_0\right]\right), &	when $|\xi|>\xi_0$.
\end{dcases*}
\end{split}
\end{equation}
Here parameters $k_1$, $k_2$, $k_3$, and $q_1$ are defined as follows
\begin{equation} \label{eq:kq-defn}
\begin{split}
k_1 &= \sqrt{\sqrt{V_0^2 + \left(1-\varepsilon \right)^2} + 1-V_0},\\
q_1 &= \sqrt{\sqrt{V_0^2 + \left(1-\varepsilon \right)^2} - 1+V_0},\\
k_2 &= \sqrt{\varepsilon}, \qquad k_3 = \sqrt{2-\varepsilon},
\end{split}
\end{equation}
with $V_0= \varkappa_0^2/2$. Eigenfunctions should satisfy the matching conditions:
\begin{equation} \label{eq:matching4u&v}
\left[\frac{u'}{u}\right]_{\xi_0}=0, \qquad \left[\frac{v'}{v}\right]_{\xi_0}=0.
\end{equation}

An explicit form of these conditions reads:
\begin{equation} \label{eq:match-varepsilon}
\begin{split}
&\frac{q_1\mu \sin q_1\xi_0 + k{_1}\sinh k{_1}\xi_0}{\mu \cos q_1\xi_0 {-} \cosh k{_1}\xi_0} = k{_3},\\
&\mu=\frac{q_1^2-V_0+\varepsilon}{k{_1}^2+V_0-\varepsilon}\times \frac{k{_1} \sinh k{_1}\xi_0 + k{_2}\cosh k{_1}\xi_0}{q_1 \sin q_1\xi_0 - k{_2}\cos q_1\xi_0}.
\end{split}
\end{equation}
The eigenvalue $\varepsilon = \varepsilon\left(V_0,\xi_0\right)$ can be found as a solution of Eqs.~\eqref{eq:match-varepsilon}.

One can make an asymptotic analysis for $\varepsilon$ when \mbox{$V_0\ll1$.} Then one can find
\begin{equation} \label{eq:energy-carbox-narrow-1}
\tag{\ref{eq:energy-carbox-narrow}$'$}
\varepsilon\approx V_0^2\xi_0^2 = \frac{\varkappa_0^4 \xi_0^2}{4}, \quad \text{when $\varkappa_0,\xi_0\ll1$}.
\end{equation}

The eigenfunctions for the scattering problem are found in the following form:
\begin{subequations} \label{eq:carbox-scat}
		\begin{equation} \label{eq:carbox-eigenfunctions-scat}
		\begin{split}
		\!\!u(\xi)\!\!&=\!\!
		\begin{dcases*}
		e^{iq\xi} + \mathcal{R} e^{-iq\xi},\!\! &when $\xi<-\xi_0$,\\
		u_1\cosh k_1\xi + u_2\sinh k_1\xi\!\!\!\\\!\!+ u_3\cos q_1\xi + u_4\sin q_1\xi,\!\! &when $|\xi|<\xi_0$,\\
		\mathcal{T} e^{ik_2\xi},\!\! &when $\xi>\xi_0$,\\
		\end{dcases*}
		\\
		\!\!v(\xi)\!\!&=\!\!
		\begin{dcases*}
		v_0 e^{k_3 (\xi+\xi_0)},\!\! &when $\xi<-\xi_0$,\\
		v_1\cosh k_1\xi + v_2\sinh k_1\xi\!\!\!\\\!\!+ v_3\cos q_1\xi + v_4\sin q_1\xi,\!\! &when $|\xi|<\xi_0$,\\
		v_5 e^{-k_3 (\xi-\xi_0)},\!\! &when $\xi>\xi_0$,
		\end{dcases*}
		\end{split}
		\end{equation}
	where $q$ is the wave number of the travelling spin wave, other parameters $k_i$ are the same as in Eqs.~\eqref{eq:kq-defn}. The usual matching conditions
	\begin{equation} \label{eq:matching-scat}
	\left[\frac{u'}{u}\right]_{\pm\xi_0}=0, \qquad \left[\frac{v'}{v}\right]_{\pm\xi_0}=0
	\end{equation}
\end{subequations}
allow us to solve the scattering problem. The resulting scattering phase is plotted in Fig.~\ref{fig:intro}(d).

\section{Wire with a sharp bend}
\label{app:sharp}
The homogeneous form of Eq.~\eqref{eq:Phi-delta} has the first integral
\begin{equation}\label{eq:first-int}
\varPhi'^2=\sin^2\varPhi+C_1.
\end{equation}
Due to the boundary conditions $\varPhi(\pm\infty)=0$ one has $C_1=0$. 
Then the solution of the model has the following form
\begin{equation} \label{eq:sharp-bend}
\tan\frac{\varPhi}{2}=
\begin{dcases*}
- e^{\xi+\xi_1},&	when $\xi<0$,\\
e^{-\xi+\xi_2}, &	when $\xi>0$.
\end{dcases*}
\end{equation}
Parameters $\xi_1$ and $\xi_2$ are determined by the matching conditions \cite{Griffiths93}
\begin{equation} \label{eq:match-cond}
\left[\varPhi'\right]_0=0,\quad \left[\varPhi\right]_0=-\beta,
\end{equation} 
which results in the solution of the form \eqref{eq:Phi-solution}.
Note that the differentiation of \eqref{eq:Phi-solution} results in $\varPhi'(\xi)=\sech(|\xi|-\xi_c)-\beta\delta(\xi)$ and as a result the potentials $V$ and $W$ in \eqref{eq:V-n-W} do not contain the $\delta$-function and have the form \eqref{eq:V-W-delta}. Since $W=0$ the EVP \eqref{eq:EVP} is split into two independent equations:
\begin{subequations} \label{eq:EVP-delta}
	\begin{align} \label{eq:u-delta}
	&-u''+\left[\varepsilon-\frac{2}{\cosh^2(|\xi|-\xi_c)}\right]u=0,\\
	\label{eq:v-delta} %
	&-v''+\left[2-\varepsilon-\frac{2}{\cosh^2(|\xi|-\xi_c)}\right]v =0.
	\end{align}
\end{subequations}
Equation \eqref{eq:u-delta} has the general solution 
\begin{equation}\label{eq:u}
\begin{split}
u=&C_1\,e^{-\sqrt{\varepsilon}|\xi|}\left[\sqrt{\varepsilon}+\tanh(|\xi|-\xi_c)\right]\\
&+C_2\,e^{\sqrt{\varepsilon}|\xi|}\left[\sqrt{\varepsilon}-\tanh(|\xi|-\xi_c)\right].
\end{split}
\end{equation}
For the case $\varOmega<1$ one has $\varepsilon>0$. In this case \eqref{eq:u-delta} with $C_2=0$ results in the bound solution \eqref{eq:uu}. It follows from \eqref{eq:u-delta} that $u'(-0)=u'(+0)$, on the other hand the solution \eqref{eq:uu} is an even function, consequently $u'(0)=0$. Applying this condition to \eqref{eq:uu} one obtains $\sqrt{\varepsilon}=f(\beta)$, where
\begin{equation}\label{eq:f}
f(\beta)=\frac{1}{2}\left[-\cos\frac{\beta}{2}+\sqrt{4-3\cos^2\frac{\beta}{2}}\right].
\end{equation}
Note that $0\le f(\beta)\le1$ if $0\le\beta\le\pi$. In the same way one obtains the condition $\sqrt{2-\varepsilon}=f(\beta)$ from \eqref{eq:v-delta}, which cannot be satisfied for $\varepsilon<1$ though. Thus only the trivial solution $v\equiv0$ of Eq.~\eqref{eq:v-delta} is possible.

For the case $\varOmega>1$  one has $\varepsilon<0$. In this case it is instructive to make a replacement $\sqrt{\varepsilon}= iq$, where  $q=\sqrt{\varOmega-1}$ has a sense of the  wave-vector. Thus, the solution of the scattering problem for  Eqs.~\eqref{eq:EVP-delta} can be presented in the form
\begin{subequations}\label{eq:scater-delta}
	\begin{align}\label{eq:u-scater}
	&u=\begin{dcases*}
	\mathcal{A}_+\left[ce^{i(q\xi+\delta_+)}+c_re^{-i(q\xi+\delta_+-\pi)}\right],& when $\xi<0$,\\
	\mathcal{A}_-c_te^{i(q\xi+\delta_-)},& when $\xi>0$,
	\end{dcases*}\\
	&v=0.
	\end{align}
\end{subequations}
Here $\mathcal{A}_\pm=\sqrt{q^2+\tanh^2(\xi\pm\xi_c)}$,  $\delta_\pm=\arg\left[iq-\tanh(\xi\pm\xi_c)\right]$ and $c$ is an arbitrary constant. Applying the conditions $u(-0)=u(+0)$ and $u'(-0)=u'(+0)$ one obtains the relations $c_t=cTe^{i\delta_t}$, $c_r=cRe^{i\delta_r}$. Since $\mathcal{A}_+(-\infty)=\mathcal{A}_-(+\infty)$, the coefficients $T$ and $R$ play the role of transmission and reflection coefficients, respectively:
\begin{subequations}\label{eq:T-R}
	\begin{align}
	&T=\frac{q \left(1+q^2\right)}{\sqrt{\cos^2\frac{\beta}{2}\sin^4\frac{\beta}{2}+q^2 \left(1+q^2\right)^2}},\\
	&R=\frac{\cos\frac{\beta}{2}\sin^2\frac{\beta}{2}}{\sqrt{\cos^2\frac{\beta}{2}\sin^4\frac{\beta}{2}+q^2 \left(1+q^2 \right)^2}}.
	\end{align}
\end{subequations} 
The corresponding phase-shifts read
\begin{subequations}\label{eq:dr-dt}
	\begin{align}\label{eq:delta_t}
	\!\!\delta_t &=\arg\left[q \left(q^2-\cos\beta \right)-i\cos\frac{\beta}{2}\left(2q^2+\sin^2\frac{\beta}{2}\right)\right]\!,\!\!\\
	\!\!\delta_r &=\delta_t+\frac{\pi}{2}.
	\end{align}
\end{subequations}
Note that for the case $\varOmega>1$ ($\varepsilon<0$) Eq.~\eqref{eq:v-delta} has only localized bound solutions. In this case $v=0$ for the same reasons as above.

%
%

\begin{thebibliography}{45}%
	\makeatletter
	\providecommand \@ifxundefined [1]{%
		\@ifx{#1\undefined}
	}%
	\providecommand \@ifnum [1]{%
		\ifnum #1\expandafter \@firstoftwo
		\else \expandafter \@secondoftwo
		\fi
	}%
	\providecommand \@ifx [1]{%
		\ifx #1\expandafter \@firstoftwo
		\else \expandafter \@secondoftwo
		\fi
	}%
	\providecommand \natexlab [1]{#1}%
	\providecommand \enquote  [1]{``#1''}%
	\providecommand \bibnamefont  [1]{#1}%
	\providecommand \bibfnamefont [1]{#1}%
	\providecommand \citenamefont [1]{#1}%
	\providecommand \href@noop [0]{\@secondoftwo}%
	\providecommand \href [0]{\begingroup \@sanitize@url \@href}%
	\providecommand \@href[1]{\@@startlink{#1}\@@href}%
	\providecommand \@@href[1]{\endgroup#1\@@endlink}%
	\providecommand \@sanitize@url [0]{\catcode `\\12\catcode `\$12\catcode
		`\&12\catcode `\#12\catcode `\^12\catcode `\_12\catcode `\%12\relax}%
	\providecommand \@@startlink[1]{}%
	\providecommand \@@endlink[0]{}%
	\providecommand \url  [0]{\begingroup\@sanitize@url \@url }%
	\providecommand \@url [1]{\endgroup\@href {#1}{\urlprefix }}%
	\providecommand \urlprefix  [0]{URL }%
	\providecommand \Eprint [0]{\href }%
	\providecommand \doibase [0]{http://dx.doi.org/}%
	\providecommand \selectlanguage [0]{\@gobble}%
	\providecommand \bibinfo  [0]{\@secondoftwo}%
	\providecommand \bibfield  [0]{\@secondoftwo}%
	\providecommand \translation [1]{[#1]}%
	\providecommand \BibitemOpen [0]{}%
	\providecommand \bibitemStop [0]{}%
	\providecommand \bibitemNoStop [0]{.\EOS\space}%
	\providecommand \EOS [0]{\spacefactor3000\relax}%
	\providecommand \BibitemShut  [1]{\csname bibitem#1\endcsname}%
	\let\auto@bib@innerbib\@empty
	\bibitem [{\citenamefont {Bloch}(1930)}]{Bloch30}%
	\BibitemOpen
	\bibfield  {author} {\bibinfo {author} {\bibfnamefont {F.}~\bibnamefont
			{Bloch}},\ }\bibfield  {title} {\enquote {\bibinfo {title} {Zur theorie des
				ferromagnetismus},}\ }\href {\doibase 10.1007/BF01339661} {\bibfield
		{journal} {\bibinfo  {journal} {Zeitschrift f{\"u}r Physik}\ }\textbf
		{\bibinfo {volume} {61}},\ \bibinfo {pages} {206--219} (\bibinfo {year}
		{1930})}\BibitemShut {NoStop}%
	\bibitem [{\citenamefont {Holstein}\ and\ \citenamefont
		{Primakoff}(1940)}]{Holstein40}%
	\BibitemOpen
	\bibfield  {author} {\bibinfo {author} {\bibfnamefont {T.}~\bibnamefont
			{Holstein}}\ and\ \bibinfo {author} {\bibfnamefont {H.}~\bibnamefont
			{Primakoff}},\ }\bibfield  {title} {\enquote {\bibinfo {title} {Field
				dependence of the intrinsic domain magnetization of a ferromagnet},}\ }\href
	{\doibase 10.1103/PhysRev.58.1098} {\bibfield  {journal} {\bibinfo  {journal}
			{Phys. Rev.}\ }\textbf {\bibinfo {volume} {58}},\ \bibinfo {pages}
		{1098--1113} (\bibinfo {year} {1940})}\BibitemShut {NoStop}%
	\bibitem [{\citenamefont {Dyson}(1956)}]{Dyson56}%
	\BibitemOpen
	\bibfield  {author} {\bibinfo {author} {\bibfnamefont {Freeman~J.}\
			\bibnamefont {Dyson}},\ }\bibfield  {title} {\enquote {\bibinfo {title}
			{General theory of spin-wave interactions},}\ }\href {\doibase
		10.1103/PhysRev.102.1217} {\bibfield  {journal} {\bibinfo  {journal} {Phys.
				Rev.}\ }\textbf {\bibinfo {volume} {102}},\ \bibinfo {pages} {1217--1230}
		(\bibinfo {year} {1956})}\BibitemShut {NoStop}%
	\bibitem [{\citenamefont {Akhiezer}\ \emph {et~al.}(1968)\citenamefont
		{Akhiezer}, \citenamefont {Bar'yakhtar},\ and\ \citenamefont
		{Peletminski\u{\i}}}]{Akhiezer68}%
	\BibitemOpen
	\bibfield  {author} {\bibinfo {author} {\bibfnamefont {A.~I.}\ \bibnamefont
			{Akhiezer}}, \bibinfo {author} {\bibfnamefont {V.~G.}\ \bibnamefont
			{Bar'yakhtar}}, \ and\ \bibinfo {author} {\bibfnamefont {S.~V.}\ \bibnamefont
			{Peletminski\u{\i}}},\ }\href@noop {} {\emph {\bibinfo {title} {Spin
				waves}}},\ edited by\ \bibinfo {editor} {\bibfnamefont {G.}~\bibnamefont
		{H{\"o}hler}}\ (\bibinfo  {publisher} {North--Holland},\ \bibinfo {address}
	{Amsterdam},\ \bibinfo {year} {1968})\BibitemShut {NoStop}%
	\bibitem [{\citenamefont {Gurevich}\ \emph {et~al.}(2000)\citenamefont
		{Gurevich}, \citenamefont {Gurevich},\ and\ \citenamefont
		{Melkov}}]{Gurevich00}%
	\BibitemOpen
	\bibfield  {author} {\bibinfo {author} {\bibfnamefont {A.~G.}\ \bibnamefont
			{Gurevich}}, \bibinfo {author} {\bibfnamefont {Alexander~G.}\ \bibnamefont
			{Gurevich}}, \ and\ \bibinfo {author} {\bibfnamefont {Gennadii~A.}\
			\bibnamefont {Melkov}},\ }\href
	{http://www.ebook.de/de/product/4297668/a_g_gurevich_alexander_g_gurevich_gennadii_a_melkov_magnetization_oscillations_and_waves.html}
	{\emph {\bibinfo {title} {Magnetization Oscillations and Waves}}}\ (\bibinfo
	{publisher} {CRC PR INC},\ \bibinfo {year} {2000})\BibitemShut {NoStop}%
	\bibitem [{\citenamefont {Kruglyak}\ \emph {et~al.}(2010)\citenamefont
		{Kruglyak}, \citenamefont {Demokritov},\ and\ \citenamefont
		{Grundler}}]{Kruglyak10a}%
	\BibitemOpen
	\bibfield  {author} {\bibinfo {author} {\bibfnamefont {V~V}\ \bibnamefont
			{Kruglyak}}, \bibinfo {author} {\bibfnamefont {S~O}\ \bibnamefont
			{Demokritov}}, \ and\ \bibinfo {author} {\bibfnamefont {D}~\bibnamefont
			{Grundler}},\ }\bibfield  {title} {\enquote {\bibinfo {title} {Magnonics},}\
	}\href {\doibase 10.1088/0022-3727/43/26/264001} {\bibfield  {journal}
		{\bibinfo  {journal} {Journal of Physics D: Applied Physics}\ }\textbf
		{\bibinfo {volume} {43}},\ \bibinfo {pages} {264001} (\bibinfo {year}
		{2010})}\BibitemShut {NoStop}%
	\bibitem [{\citenamefont {Lenk}\ \emph {et~al.}(2011)\citenamefont {Lenk},
		\citenamefont {Ulrichs}, \citenamefont {Garbs},\ and\ \citenamefont
		{M\"unzenberg}}]{Lenk11}%
	\BibitemOpen
	\bibfield  {author} {\bibinfo {author} {\bibfnamefont {B.}~\bibnamefont
			{Lenk}}, \bibinfo {author} {\bibfnamefont {H.}~\bibnamefont {Ulrichs}},
		\bibinfo {author} {\bibfnamefont {F.}~\bibnamefont {Garbs}}, \ and\ \bibinfo
		{author} {\bibfnamefont {M.}~\bibnamefont {M\"unzenberg}},\ }\bibfield
	{title} {\enquote {\bibinfo {title} {The building blocks of magnonics},}\
	}\href {\doibase 10.1016/j.physrep.2011.06.003} {\bibfield  {journal}
		{\bibinfo  {journal} {Physics Reports}\ }\textbf {\bibinfo {volume} {507}},\
		\bibinfo {pages} {107--136} (\bibinfo {year} {2011})}\BibitemShut {NoStop}%
	\bibitem [{\citenamefont {Chumak}\ \emph {et~al.}(2014)\citenamefont {Chumak},
		\citenamefont {Serga},\ and\ \citenamefont {Hillebrands}}]{Chumak14}%
	\BibitemOpen
	\bibfield  {author} {\bibinfo {author} {\bibfnamefont {Andrii~V.}\
			\bibnamefont {Chumak}}, \bibinfo {author} {\bibfnamefont {Alexander~A.}\
			\bibnamefont {Serga}}, \ and\ \bibinfo {author} {\bibfnamefont {Burkard}\
			\bibnamefont {Hillebrands}},\ }\bibfield  {title} {\enquote {\bibinfo {title}
			{Magnon transistor for all-magnon data processing},}\ }\href {\doibase
		10.1038/ncomms5700} {\bibfield  {journal} {\bibinfo  {journal} {Nature
				Communications}\ }\textbf {\bibinfo {volume} {5}},\ \bibinfo {pages} {4700}
		(\bibinfo {year} {2014})}\BibitemShut {NoStop}%
	\bibitem [{\citenamefont {Chumak}\ \emph {et~al.}(2015)\citenamefont {Chumak},
		\citenamefont {Vasyuchka}, \citenamefont {Serga},\ and\ \citenamefont
		{Hillebrands}}]{Chumak15}%
	\BibitemOpen
	\bibfield  {author} {\bibinfo {author} {\bibfnamefont {A.~V.}\ \bibnamefont
			{Chumak}}, \bibinfo {author} {\bibfnamefont {V.~I.}\ \bibnamefont
			{Vasyuchka}}, \bibinfo {author} {\bibfnamefont {A.~A.}\ \bibnamefont
			{Serga}}, \ and\ \bibinfo {author} {\bibfnamefont {B.}~\bibnamefont
			{Hillebrands}},\ }\bibfield  {title} {\enquote {\bibinfo {title} {Magnon
				spintronics},}\ }\href {\doibase 10.1038/nphys3347} {\bibfield  {journal}
		{\bibinfo  {journal} {Nat Phys}\ }\textbf {\bibinfo {volume} {11}},\ \bibinfo
		{pages} {453--461} (\bibinfo {year} {2015})}\BibitemShut {NoStop}%
	\bibitem [{\citenamefont {Vogt}\ \emph {et~al.}(2012)\citenamefont {Vogt},
		\citenamefont {Schultheiss}, \citenamefont {Jain}, \citenamefont {Pearson},
		\citenamefont {Hoffmann}, \citenamefont {Bader},\ and\ \citenamefont
		{Hillebrands}}]{Vogt12}%
	\BibitemOpen
	\bibfield  {author} {\bibinfo {author} {\bibfnamefont {K.}~\bibnamefont
			{Vogt}}, \bibinfo {author} {\bibfnamefont {H.}~\bibnamefont {Schultheiss}},
		\bibinfo {author} {\bibfnamefont {S.}~\bibnamefont {Jain}}, \bibinfo {author}
		{\bibfnamefont {J.~E.}\ \bibnamefont {Pearson}}, \bibinfo {author}
		{\bibfnamefont {A.}~\bibnamefont {Hoffmann}}, \bibinfo {author}
		{\bibfnamefont {S.~D.}\ \bibnamefont {Bader}}, \ and\ \bibinfo {author}
		{\bibfnamefont {B.}~\bibnamefont {Hillebrands}},\ }\bibfield  {title}
	{\enquote {\bibinfo {title} {Spin waves turning a corner},}\ }\href {\doibase
		10.1063/1.4738887} {\bibfield  {journal} {\bibinfo  {journal} {Applied
				Physics Letters}\ }\textbf {\bibinfo {volume} {101}},\ \bibinfo {pages}
		{042410} (\bibinfo {year} {2012})}\BibitemShut {NoStop}%
	\bibitem [{\citenamefont {Xing}\ \emph {et~al.}(2013)\citenamefont {Xing},
		\citenamefont {Yu}, \citenamefont {Li},\ and\ \citenamefont
		{Huang}}]{Xing13}%
	\BibitemOpen
	\bibfield  {author} {\bibinfo {author} {\bibfnamefont {Xiangjun}\
			\bibnamefont {Xing}}, \bibinfo {author} {\bibfnamefont {Yongli}\ \bibnamefont
			{Yu}}, \bibinfo {author} {\bibfnamefont {Shuwei}\ \bibnamefont {Li}}, \ and\
		\bibinfo {author} {\bibfnamefont {Xiaohong}\ \bibnamefont {Huang}},\
	}\bibfield  {title} {\enquote {\bibinfo {title} {How do spin waves pass
				through a bend?}}\ }\href {\doibase 10.1038/srep02958} {\bibfield  {journal}
		{\bibinfo  {journal} {Scientific Reports}\ }\textbf {\bibinfo {volume} {3}}
		(\bibinfo {year} {2013}),\ 10.1038/srep02958}\BibitemShut {NoStop}%
	\bibitem [{\citenamefont {Haldar}\ \emph {et~al.}(2016)\citenamefont {Haldar},
		\citenamefont {Kumar},\ and\ \citenamefont {Adeyeye}}]{Haldar16}%
	\BibitemOpen
	\bibfield  {author} {\bibinfo {author} {\bibfnamefont {Arabinda}\
			\bibnamefont {Haldar}}, \bibinfo {author} {\bibfnamefont {Dheeraj}\
			\bibnamefont {Kumar}}, \ and\ \bibinfo {author} {\bibfnamefont
			{Adekunle~Olusola}\ \bibnamefont {Adeyeye}},\ }\bibfield  {title} {\enquote
		{\bibinfo {title} {A reconfigurable waveguide for energy-efficient
				transmission and local manipulation of information in a nanomagnetic
				device},}\ }\href {\doibase 10.1038/nnano.2015.332} {\bibfield  {journal}
		{\bibinfo  {journal} {Nature Nanotechnology}\ }\textbf {\bibinfo {volume}
			{11}},\ \bibinfo {pages} {437--443} (\bibinfo {year} {2016})}\BibitemShut
	{NoStop}%
	\bibitem [{\citenamefont {P{\'e}rez}\ \emph {et~al.}(2015)\citenamefont
		{P{\'e}rez}, \citenamefont {Melzer}, \citenamefont {Makarov}, \citenamefont
		{Uebersch{\"a}r}, \citenamefont {Ecke}, \citenamefont {Schulz},\ and\
		\citenamefont {Schmidt}}]{Perez15}%
	\BibitemOpen
	\bibfield  {author} {\bibinfo {author} {\bibfnamefont {Nicol{\'a}s}\
			\bibnamefont {P{\'e}rez}}, \bibinfo {author} {\bibfnamefont {Michael}\
			\bibnamefont {Melzer}}, \bibinfo {author} {\bibfnamefont {Denys}\
			\bibnamefont {Makarov}}, \bibinfo {author} {\bibfnamefont {Olaf}\
			\bibnamefont {Uebersch{\"a}r}}, \bibinfo {author} {\bibfnamefont {Ramona}\
			\bibnamefont {Ecke}}, \bibinfo {author} {\bibfnamefont {Stefan~E.}\
			\bibnamefont {Schulz}}, \ and\ \bibinfo {author} {\bibfnamefont {Oliver~G.}\
			\bibnamefont {Schmidt}},\ }\bibfield  {title} {\enquote {\bibinfo {title}
			{High-performance giant magnetoresistive sensorics on flexible si
				membranes},}\ }\href {\doibase 10.1063/1.4918652} {\bibfield  {journal}
		{\bibinfo  {journal} {Appl. Phys. Lett.}\ }\textbf {\bibinfo {volume}
			{106}},\ \bibinfo {pages} {153501} (\bibinfo {year} {2015})}\BibitemShut
	{NoStop}%
	\bibitem [{\citenamefont {Melzer}\ \emph {et~al.}(2011)\citenamefont {Melzer},
		\citenamefont {Makarov}, \citenamefont {Calvimontes}, \citenamefont
		{Karnaushenko}, \citenamefont {Baunack}, \citenamefont {Kaltofen},
		\citenamefont {Mei},\ and\ \citenamefont {Schmidt}}]{Melzer11}%
	\BibitemOpen
	\bibfield  {author} {\bibinfo {author} {\bibfnamefont {Michael}\ \bibnamefont
			{Melzer}}, \bibinfo {author} {\bibfnamefont {Denys}\ \bibnamefont {Makarov}},
		\bibinfo {author} {\bibfnamefont {Alfredo}\ \bibnamefont {Calvimontes}},
		\bibinfo {author} {\bibfnamefont {Daniil}\ \bibnamefont {Karnaushenko}},
		\bibinfo {author} {\bibfnamefont {Stefan}\ \bibnamefont {Baunack}}, \bibinfo
		{author} {\bibfnamefont {Rainer}\ \bibnamefont {Kaltofen}}, \bibinfo {author}
		{\bibfnamefont {Yongfeng}\ \bibnamefont {Mei}}, \ and\ \bibinfo {author}
		{\bibfnamefont {Oliver~G.}\ \bibnamefont {Schmidt}},\ }\bibfield  {title}
	{\enquote {\bibinfo {title} {Stretchable magnetoelectronics},}\ }\href
	{\doibase 10.1021/nl201108b} {\bibfield  {journal} {\bibinfo  {journal} {Nano
				Letters}\ }\textbf {\bibinfo {volume} {11}},\ \bibinfo {pages} {2522--2526}
		(\bibinfo {year} {2011})},\ \Eprint
	{http://arxiv.org/abs/http://pubs.acs.org/doi/pdf/10.1021/nl201108b}
	{http://pubs.acs.org/doi/pdf/10.1021/nl201108b} \BibitemShut {NoStop}%
	\bibitem [{\citenamefont {Makarov}\ \emph {et~al.}(2013)\citenamefont
		{Makarov}, \citenamefont {Karnaushenko},\ and\ \citenamefont
		{Schmidt}}]{Makarov13b}%
	\BibitemOpen
	\bibfield  {author} {\bibinfo {author} {\bibfnamefont {Denys}\ \bibnamefont
			{Makarov}}, \bibinfo {author} {\bibfnamefont {Daniil}\ \bibnamefont
			{Karnaushenko}}, \ and\ \bibinfo {author} {\bibfnamefont {Oliver~G.}\
			\bibnamefont {Schmidt}},\ }\bibfield  {title} {\enquote {\bibinfo {title}
			{Printable magnetoelectronics},}\ }\href {\doibase 10.1002/cphc.201300162}
	{\bibfield  {journal} {\bibinfo  {journal} {ChemPhysChem}\ }\textbf {\bibinfo
			{volume} {14}},\ \bibinfo {pages} {1771--1776} (\bibinfo {year}
		{2013})}\BibitemShut {NoStop}%
	\bibitem [{\citenamefont {Streubel}\ \emph {et~al.}(2016)\citenamefont
		{Streubel}, \citenamefont {Fischer}, \citenamefont {Kronast}, \citenamefont
		{Kravchuk}, \citenamefont {Sheka}, \citenamefont {Gaididei}, \citenamefont
		{Schmidt},\ and\ \citenamefont {Makarov}}]{Streubel16a}%
	\BibitemOpen
	\bibfield  {author} {\bibinfo {author} {\bibfnamefont {Robert}\ \bibnamefont
			{Streubel}}, \bibinfo {author} {\bibfnamefont {Peter}\ \bibnamefont
			{Fischer}}, \bibinfo {author} {\bibfnamefont {Florian}\ \bibnamefont
			{Kronast}}, \bibinfo {author} {\bibfnamefont {Volodymyr~P.}\ \bibnamefont
			{Kravchuk}}, \bibinfo {author} {\bibfnamefont {Denis~D.}\ \bibnamefont
			{Sheka}}, \bibinfo {author} {\bibfnamefont {Yuri}\ \bibnamefont {Gaididei}},
		\bibinfo {author} {\bibfnamefont {Oliver~G.}\ \bibnamefont {Schmidt}}, \ and\
		\bibinfo {author} {\bibfnamefont {Denys}\ \bibnamefont {Makarov}},\
	}\bibfield  {title} {\enquote {\bibinfo {title} {Magnetism in curved
				geometries (topical review)},}\ }\href {\doibase
		10.1088/0022-3727/49/36/363001} {\bibfield  {journal} {\bibinfo  {journal}
			{Journal of Physics D: Applied Physics}\ }\textbf {\bibinfo {volume} {49}},\
		\bibinfo {pages} {363001} (\bibinfo {year} {2016})}\BibitemShut {NoStop}%
	\bibitem [{\citenamefont {Gaididei}\ \emph {et~al.}(2014)\citenamefont
		{Gaididei}, \citenamefont {Kravchuk},\ and\ \citenamefont
		{Sheka}}]{Gaididei14}%
	\BibitemOpen
	\bibfield  {author} {\bibinfo {author} {\bibfnamefont {Yuri}\ \bibnamefont
			{Gaididei}}, \bibinfo {author} {\bibfnamefont {Volodymyr~P.}\ \bibnamefont
			{Kravchuk}}, \ and\ \bibinfo {author} {\bibfnamefont {Denis~D.}\ \bibnamefont
			{Sheka}},\ }\bibfield  {title} {\enquote {\bibinfo {title} {Curvature effects
				in thin magnetic shells},}\ }\href {\doibase 10.1103/PhysRevLett.112.257203}
	{\bibfield  {journal} {\bibinfo  {journal} {Phys. Rev. Lett.}\ }\textbf
		{\bibinfo {volume} {112}},\ \bibinfo {pages} {257203} (\bibinfo {year}
		{2014})}\BibitemShut {NoStop}%
	\bibitem [{\citenamefont {Sheka}\ \emph
		{et~al.}(2015{\natexlab{a}})\citenamefont {Sheka}, \citenamefont {Kravchuk},\
		and\ \citenamefont {Gaididei}}]{Sheka15}%
	\BibitemOpen
	\bibfield  {author} {\bibinfo {author} {\bibfnamefont {Denis~D.}\
			\bibnamefont {Sheka}}, \bibinfo {author} {\bibfnamefont {Volodymyr~P.}\
			\bibnamefont {Kravchuk}}, \ and\ \bibinfo {author} {\bibfnamefont {Yuri}\
			\bibnamefont {Gaididei}},\ }\bibfield  {title} {\enquote {\bibinfo {title}
			{Curvature effects in statics and dynamics of low dimensional magnets},}\
	}\href {\doibase http://dx.doi.org/10.1088/1751-8113/48/12/125202} {\bibfield
		{journal} {\bibinfo  {journal} {Journal of Physics A: Mathematical and
				Theoretical}\ }\textbf {\bibinfo {volume} {48}},\ \bibinfo {pages} {125202}
		(\bibinfo {year} {2015}{\natexlab{a}})}\BibitemShut {NoStop}%
	\bibitem [{\citenamefont {Gaididei}\ \emph {et~al.}(2017)\citenamefont
		{Gaididei}, \citenamefont {Goussev}, \citenamefont {Kravchuk}, \citenamefont
		{Pylypovskyi}, \citenamefont {Robbins}, \citenamefont {Sheka}, \citenamefont
		{Slastikov},\ and\ \citenamefont {Vasylkevych}}]{Gaididei17a}%
	\BibitemOpen
	\bibfield  {author} {\bibinfo {author} {\bibfnamefont {Yuri~B}\ \bibnamefont
			{Gaididei}}, \bibinfo {author} {\bibfnamefont {Arseni}\ \bibnamefont
			{Goussev}}, \bibinfo {author} {\bibfnamefont {Volodymyr~P}\ \bibnamefont
			{Kravchuk}}, \bibinfo {author} {\bibfnamefont {Oleksandr~V}\ \bibnamefont
			{Pylypovskyi}}, \bibinfo {author} {\bibfnamefont {Jonathan~M}\ \bibnamefont
			{Robbins}}, \bibinfo {author} {\bibfnamefont {Denis}\ \bibnamefont {Sheka}},
		\bibinfo {author} {\bibfnamefont {Valeriy}\ \bibnamefont {Slastikov}}, \ and\
		\bibinfo {author} {\bibfnamefont {Sergiy}\ \bibnamefont {Vasylkevych}},\
	}\bibfield  {title} {\enquote {\bibinfo {title} {Magnetization in narrow
				ribbons: curvature effects},}\ }\href {\doibase 10.1088/1751-8121/aa8179}
	{\bibfield  {journal} {\bibinfo  {journal} {Journal of Physics A:
				Mathematical and Theoretical}\ }\textbf {\bibinfo {volume} {50}},\ \bibinfo
		{pages} {385401} (\bibinfo {year} {2017})}\BibitemShut {NoStop}%
	\bibitem [{\citenamefont {Hertel}(2013)}]{Hertel13a}%
	\BibitemOpen
	\bibfield  {author} {\bibinfo {author} {\bibfnamefont {Riccardo}\
			\bibnamefont {Hertel}},\ }\bibfield  {title} {\enquote {\bibinfo {title}
			{Curvature--induced magnetochirality},}\ }\href {\doibase
		10.1142/s2010324713400092} {\bibfield  {journal} {\bibinfo  {journal} {SPIN}\
		}\textbf {\bibinfo {volume} {03}},\ \bibinfo {pages} {1340009} (\bibinfo
		{year} {2013})}\BibitemShut {NoStop}%
	\bibitem [{\citenamefont {Yershov}\ \emph {et~al.}(2015)\citenamefont
		{Yershov}, \citenamefont {Kravchuk}, \citenamefont {Sheka},\ and\
		\citenamefont {Gaididei}}]{Yershov15b}%
	\BibitemOpen
	\bibfield  {author} {\bibinfo {author} {\bibfnamefont {Kostiantyn~V.}\
			\bibnamefont {Yershov}}, \bibinfo {author} {\bibfnamefont {Volodymyr~P.}\
			\bibnamefont {Kravchuk}}, \bibinfo {author} {\bibfnamefont {Denis~D.}\
			\bibnamefont {Sheka}}, \ and\ \bibinfo {author} {\bibfnamefont {Yuri}\
			\bibnamefont {Gaididei}},\ }\bibfield  {title} {\enquote {\bibinfo {title}
			{Curvature-induced domain wall pinning},}\ }\href {\doibase
		10.1103/PhysRevB.92.104412} {\bibfield  {journal} {\bibinfo  {journal} {Phys.
				Rev. B}\ }\textbf {\bibinfo {volume} {92}},\ \bibinfo {pages} {104412}
		(\bibinfo {year} {2015})}\BibitemShut {NoStop}%
	\bibitem [{\citenamefont {Pylypovskyi}\ \emph {et~al.}(2015)\citenamefont
		{Pylypovskyi}, \citenamefont {Kravchuk}, \citenamefont {Sheka}, \citenamefont
		{Makarov}, \citenamefont {Schmidt},\ and\ \citenamefont
		{Gaididei}}]{Pylypovskyi15b}%
	\BibitemOpen
	\bibfield  {author} {\bibinfo {author} {\bibfnamefont {Oleksandr~V.}\
			\bibnamefont {Pylypovskyi}}, \bibinfo {author} {\bibfnamefont {Volodymyr~P.}\
			\bibnamefont {Kravchuk}}, \bibinfo {author} {\bibfnamefont {Denis~D.}\
			\bibnamefont {Sheka}}, \bibinfo {author} {\bibfnamefont {Denys}\ \bibnamefont
			{Makarov}}, \bibinfo {author} {\bibfnamefont {Oliver~G.}\ \bibnamefont
			{Schmidt}}, \ and\ \bibinfo {author} {\bibfnamefont {Yuri}\ \bibnamefont
			{Gaididei}},\ }\bibfield  {title} {\enquote {\bibinfo {title} {Coupling of
				chiralities in spin and physical spaces: {T}he {M}\"obius ring as a case
				study},}\ }\href {\doibase 10.1103/PhysRevLett.114.197204} {\bibfield
		{journal} {\bibinfo  {journal} {Phys. Rev. Lett.}\ }\textbf {\bibinfo
			{volume} {114}},\ \bibinfo {pages} {197204} (\bibinfo {year}
		{2015})}\BibitemShut {NoStop}%
	\bibitem [{\citenamefont {Sheka}\ \emph
		{et~al.}(2015{\natexlab{b}})\citenamefont {Sheka}, \citenamefont {Kravchuk},
		\citenamefont {Yershov},\ and\ \citenamefont {Gaididei}}]{Sheka15c}%
	\BibitemOpen
	\bibfield  {author} {\bibinfo {author} {\bibfnamefont {Denis~D.}\
			\bibnamefont {Sheka}}, \bibinfo {author} {\bibfnamefont {Volodymyr~P.}\
			\bibnamefont {Kravchuk}}, \bibinfo {author} {\bibfnamefont {Kostiantyn~V.}\
			\bibnamefont {Yershov}}, \ and\ \bibinfo {author} {\bibfnamefont {Yuri}\
			\bibnamefont {Gaididei}},\ }\bibfield  {title} {\enquote {\bibinfo {title}
			{Torsion-induced effects in magnetic nanowires},}\ }\href {\doibase
		10.1103/PhysRevB.92.054417} {\bibfield  {journal} {\bibinfo  {journal} {Phys.
				Rev. B}\ }\textbf {\bibinfo {volume} {92}},\ \bibinfo {pages} {054417}
		(\bibinfo {year} {2015}{\natexlab{b}})}\BibitemShut {NoStop}%
	\bibitem [{\citenamefont {Yershov}\ \emph {et~al.}(2016)\citenamefont
		{Yershov}, \citenamefont {Kravchuk}, \citenamefont {Sheka},\ and\
		\citenamefont {Gaididei}}]{Yershov16}%
	\BibitemOpen
	\bibfield  {author} {\bibinfo {author} {\bibfnamefont {Kostiantyn~V.}\
			\bibnamefont {Yershov}}, \bibinfo {author} {\bibfnamefont {Volodymyr~P.}\
			\bibnamefont {Kravchuk}}, \bibinfo {author} {\bibfnamefont {Denis~D.}\
			\bibnamefont {Sheka}}, \ and\ \bibinfo {author} {\bibfnamefont {Yuri}\
			\bibnamefont {Gaididei}},\ }\bibfield  {title} {\enquote {\bibinfo {title}
			{Curvature and torsion effects in spin-current driven domain wall motion},}\
	}\href {\doibase 10.1103/PhysRevB.93.094418} {\bibfield  {journal} {\bibinfo
			{journal} {Phys. Rev. B}\ }\textbf {\bibinfo {volume} {93}},\ \bibinfo
		{pages} {094418} (\bibinfo {year} {2016})}\BibitemShut {NoStop}%
	\bibitem [{\citenamefont {Pylypovskyi}\ \emph {et~al.}(2016)\citenamefont
		{Pylypovskyi}, \citenamefont {Sheka}, \citenamefont {Kravchuk}, \citenamefont
		{Yershov}, \citenamefont {Makarov},\ and\ \citenamefont
		{Gaididei}}]{Pylypovskyi16}%
	\BibitemOpen
	\bibfield  {author} {\bibinfo {author} {\bibfnamefont {Oleksandr~V.}\
			\bibnamefont {Pylypovskyi}}, \bibinfo {author} {\bibfnamefont {Denis~D.}\
			\bibnamefont {Sheka}}, \bibinfo {author} {\bibfnamefont {Volodymyr~P.}\
			\bibnamefont {Kravchuk}}, \bibinfo {author} {\bibfnamefont {Kostiantyn~V.}\
			\bibnamefont {Yershov}}, \bibinfo {author} {\bibfnamefont {Denys}\
			\bibnamefont {Makarov}}, \ and\ \bibinfo {author} {\bibfnamefont {Yuri}\
			\bibnamefont {Gaididei}},\ }\bibfield  {title} {\enquote {\bibinfo {title}
			{Rashba torque driven domain wall motion in magnetic helices},}\ }\href
	{\doibase 10.1038/srep23316} {\bibfield  {journal} {\bibinfo  {journal}
			{Scientific Reports}\ }\textbf {\bibinfo {volume} {6}},\ \bibinfo {pages}
		{23316} (\bibinfo {year} {2016})}\BibitemShut {NoStop}%
	\bibitem [{\citenamefont {Volkov}\ \emph {et~al.}(2018)\citenamefont {Volkov},
		\citenamefont {Sheka}, \citenamefont {Gaididei}, \citenamefont {Kravchuk},
		\citenamefont {R\"o{\ss}ler}, \citenamefont {Fassbender},\ and\ \citenamefont
		{Makarov}}]{Volkov18}%
	\BibitemOpen
	\bibfield  {author} {\bibinfo {author} {\bibfnamefont {Oleksii~M.}\
			\bibnamefont {Volkov}}, \bibinfo {author} {\bibfnamefont {Denis~D.}\
			\bibnamefont {Sheka}}, \bibinfo {author} {\bibfnamefont {Yuri}\ \bibnamefont
			{Gaididei}}, \bibinfo {author} {\bibfnamefont {Volodymyr~P.}\ \bibnamefont
			{Kravchuk}}, \bibinfo {author} {\bibfnamefont {Ulrich~K.}\ \bibnamefont
			{R\"o{\ss}ler}}, \bibinfo {author} {\bibfnamefont {J\"urgen}\ \bibnamefont
			{Fassbender}}, \ and\ \bibinfo {author} {\bibfnamefont {Denys}\ \bibnamefont
			{Makarov}},\ }\bibfield  {title} {\enquote {\bibinfo {title} {Mesoscale
				dzyaloshinskii-moriya interaction: geometrical tailoring of the
				magnetochirality},}\ }\href {\doibase 10.1038/s41598-017-18835-4} {\bibfield
		{journal} {\bibinfo  {journal} {Scientific Reports}\ }\textbf {\bibinfo
			{volume} {8}},\ \bibinfo {pages} {866} (\bibinfo {year} {2018})}\BibitemShut
	{NoStop}%
	\bibitem [{\citenamefont {Kravchuk}\ \emph {et~al.}(2016)\citenamefont
		{Kravchuk}, \citenamefont {R\"o\ss{}ler}, \citenamefont {Volkov},
		\citenamefont {Sheka}, \citenamefont {van~den Brink}, \citenamefont
		{Makarov}, \citenamefont {Fuchs}, \citenamefont {Fangohr},\ and\
		\citenamefont {Gaididei}}]{Kravchuk16a}%
	\BibitemOpen
	\bibfield  {author} {\bibinfo {author} {\bibfnamefont {Volodymyr~P.}\
			\bibnamefont {Kravchuk}}, \bibinfo {author} {\bibfnamefont {Ulrich~K.}\
			\bibnamefont {R\"o\ss{}ler}}, \bibinfo {author} {\bibfnamefont {Oleksii~M.}\
			\bibnamefont {Volkov}}, \bibinfo {author} {\bibfnamefont {Denis~D.}\
			\bibnamefont {Sheka}}, \bibinfo {author} {\bibfnamefont {Jeroen}\
			\bibnamefont {van~den Brink}}, \bibinfo {author} {\bibfnamefont {Denys}\
			\bibnamefont {Makarov}}, \bibinfo {author} {\bibfnamefont {Hagen}\
			\bibnamefont {Fuchs}}, \bibinfo {author} {\bibfnamefont {Hans}\ \bibnamefont
			{Fangohr}}, \ and\ \bibinfo {author} {\bibfnamefont {Yuri}\ \bibnamefont
			{Gaididei}},\ }\bibfield  {title} {\enquote {\bibinfo {title} {Topologically
				stable magnetization states on a spherical shell: Curvature-stabilized
				skyrmions},}\ }\href {\doibase 10.1103/PhysRevB.94.144402} {\bibfield
		{journal} {\bibinfo  {journal} {Phys. Rev. B}\ }\textbf {\bibinfo {volume}
			{94}},\ \bibinfo {pages} {144402} (\bibinfo {year} {2016})}\BibitemShut
	{NoStop}%
	\bibitem [{\citenamefont {Kravchuk}\ \emph
		{et~al.}(2018{\natexlab{a}})\citenamefont {Kravchuk}, \citenamefont {Sheka},
		\citenamefont {K\'akay}, \citenamefont {Volkov}, \citenamefont
		{R\"o\ss{}ler}, \citenamefont {van~den Brink}, \citenamefont {Makarov},\ and\
		\citenamefont {Gaididei}}]{Kravchuk18a}%
	\BibitemOpen
	\bibfield  {author} {\bibinfo {author} {\bibfnamefont {Volodymyr~P.}\
			\bibnamefont {Kravchuk}}, \bibinfo {author} {\bibfnamefont {Denis~D.}\
			\bibnamefont {Sheka}}, \bibinfo {author} {\bibfnamefont {Attila}\
			\bibnamefont {K\'akay}}, \bibinfo {author} {\bibfnamefont {Oleksii~M.}\
			\bibnamefont {Volkov}}, \bibinfo {author} {\bibfnamefont {Ulrich~K.}\
			\bibnamefont {R\"o\ss{}ler}}, \bibinfo {author} {\bibfnamefont {Jeroen}\
			\bibnamefont {van~den Brink}}, \bibinfo {author} {\bibfnamefont {Denys}\
			\bibnamefont {Makarov}}, \ and\ \bibinfo {author} {\bibfnamefont {Yuri}\
			\bibnamefont {Gaididei}},\ }\bibfield  {title} {\enquote {\bibinfo {title}
			{Multiplet of skyrmion states on a curvilinear defect: Reconfigurable
				skyrmion lattices},}\ }\href {\doibase 10.1103/PhysRevLett.120.067201}
	{\bibfield  {journal} {\bibinfo  {journal} {Phys. Rev. Lett.}\ }\textbf
		{\bibinfo {volume} {120}},\ \bibinfo {pages} {067201} (\bibinfo {year}
		{2018}{\natexlab{a}})}\BibitemShut {NoStop}%
	\bibitem [{\citenamefont {Slastikov}\ and\ \citenamefont
		{Sonnenberg}(2012)}]{Slastikov12}%
	\BibitemOpen
	\bibfield  {author} {\bibinfo {author} {\bibfnamefont {V.~V.}\ \bibnamefont
			{Slastikov}}\ and\ \bibinfo {author} {\bibfnamefont {C.}~\bibnamefont
			{Sonnenberg}},\ }\bibfield  {title} {\enquote {\bibinfo {title} {Reduced
				models for ferromagnetic nanowires},}\ }\href {\doibase
		10.1093/imamat/hxr019} {\bibfield  {journal} {\bibinfo  {journal} {IMA
				Journal of Applied Mathematics}\ }\textbf {\bibinfo {volume} {77}},\ \bibinfo
		{pages} {220--235} (\bibinfo {year} {2012})}\BibitemShut {NoStop}%
	\bibitem [{\citenamefont {Sheka}\ \emph {et~al.}(2004)\citenamefont {Sheka},
		\citenamefont {Yastremsky}, \citenamefont {Ivanov}, \citenamefont {Wysin},\
		and\ \citenamefont {Mertens}}]{Sheka04}%
	\BibitemOpen
	\bibfield  {author} {\bibinfo {author} {\bibfnamefont {Denis~D.}\
			\bibnamefont {Sheka}}, \bibinfo {author} {\bibfnamefont {Ivan~A.}\
			\bibnamefont {Yastremsky}}, \bibinfo {author} {\bibfnamefont {Boris~A.}\
			\bibnamefont {Ivanov}}, \bibinfo {author} {\bibfnamefont {Gary~M.}\
			\bibnamefont {Wysin}}, \ and\ \bibinfo {author} {\bibfnamefont {Franz~G.}\
			\bibnamefont {Mertens}},\ }\bibfield  {title} {\enquote {\bibinfo {title}
			{Amplitudes for magnon scattering by vortices in two--dimensional weakly
				easy--plane ferromagnets},}\ }\href
	{http://link.aps.org/abstract/PRB/v69/e054429} {\bibfield  {journal}
		{\bibinfo  {journal} {Phys. Rev. B}\ }\textbf {\bibinfo {volume} {69}},\
		\bibinfo {eid} {054429} (\bibinfo {year} {2004})}\BibitemShut {NoStop}%
	\bibitem [{\citenamefont {Ivanov}\ and\ \citenamefont
		{Sheka}(2005)}]{Ivanov05b}%
	\BibitemOpen
	\bibfield  {author} {\bibinfo {author} {\bibfnamefont {B.~A.}\ \bibnamefont
			{Ivanov}}\ and\ \bibinfo {author} {\bibfnamefont {D.~D.}\ \bibnamefont
			{Sheka}},\ }\bibfield  {title} {\enquote {\bibinfo {title} {Local magnon
				modes and the dynamics of a small--radius two--dimensional magnetic soliton
				in an easy-axis ferromagnet},}\ }\href {\doibase 10.1134/1.2142872}
	{\bibfield  {journal} {\bibinfo  {journal} {JETP Lett.}\ }\textbf {\bibinfo
			{volume} {82}},\ \bibinfo {pages} {436--440} (\bibinfo {year}
		{2005})}\BibitemShut {NoStop}%
	\bibitem [{\citenamefont {Kravchuk}\ \emph
		{et~al.}(2018{\natexlab{b}})\citenamefont {Kravchuk}, \citenamefont {Sheka},
		\citenamefont {R\"o\ss{}ler}, \citenamefont {van~den Brink},\ and\
		\citenamefont {Gaididei}}]{Kravchuk18}%
	\BibitemOpen
	\bibfield  {author} {\bibinfo {author} {\bibfnamefont {Volodymyr~P.}\
			\bibnamefont {Kravchuk}}, \bibinfo {author} {\bibfnamefont {Denis~D.}\
			\bibnamefont {Sheka}}, \bibinfo {author} {\bibfnamefont {Ulrich~K.}\
			\bibnamefont {R\"o\ss{}ler}}, \bibinfo {author} {\bibfnamefont {Jeroen}\
			\bibnamefont {van~den Brink}}, \ and\ \bibinfo {author} {\bibfnamefont
			{Yuri}\ \bibnamefont {Gaididei}},\ }\bibfield  {title} {\enquote {\bibinfo
			{title} {Spin eigenmodes of magnetic skyrmions and the problem of the
				effective skyrmion mass},}\ }\href {\doibase 10.1103/PhysRevB.97.064403}
	{\bibfield  {journal} {\bibinfo  {journal} {Phys. Rev. B}\ }\textbf {\bibinfo
			{volume} {97}},\ \bibinfo {pages} {064403} (\bibinfo {year}
		{2018}{\natexlab{b}})}\BibitemShut {NoStop}%
	\bibitem [{\citenamefont {Simon}(1976)}]{Simon76}%
	\BibitemOpen
	\bibfield  {author} {\bibinfo {author} {\bibfnamefont {Barry}\ \bibnamefont
			{Simon}},\ }\bibfield  {title} {\enquote {\bibinfo {title} {The bound state
				of weakly coupled schr{\"o}dinger operators in one and two dimensions},}\
	}\href {\doibase 10.1016/0003-4916(76)90038-5} {\bibfield  {journal}
		{\bibinfo  {journal} {Annals of Physics}\ }\textbf {\bibinfo {volume} {97}},\
		\bibinfo {pages} {279--288} (\bibinfo {year} {1976})}\BibitemShut {NoStop}%
	\bibitem [{\citenamefont {Landau}\ and\ \citenamefont
		{Lifshitz}(1999)}]{LandauIII}%
	\BibitemOpen
	\bibfield  {author} {\bibinfo {author} {\bibfnamefont {L.~D.}\ \bibnamefont
			{Landau}}\ and\ \bibinfo {author} {\bibfnamefont {E.~M.}\ \bibnamefont
			{Lifshitz}},\ }\href@noop {} {\emph {\bibinfo {title} {Quantum mechanics ---
				non--relativistic theory}}}\ (\bibinfo  {publisher}
	{Butterworth--Heinemann},\ \bibinfo {address} {Linacre House, Jordan Hill,
		Oxford},\ \bibinfo {year} {1999})\BibitemShut {NoStop}%
	\bibitem [{\citenamefont {Swan}(1963)}]{Swan63}%
	\BibitemOpen
	\bibfield  {author} {\bibinfo {author} {\bibfnamefont {P.}~\bibnamefont
			{Swan}},\ }\bibfield  {title} {\enquote {\bibinfo {title} {Asymptotic
				phase-shifts and bound states for two-body central interactions},}\ }\href
	{http://www.sciencedirect.com/science/article/B73DR-470WBG8-13R/2/0c08b94e655b4b11e41b1300e638d692}
	{\bibfield  {journal} {\bibinfo  {journal} {Nuclear Physics}\ }\textbf
		{\bibinfo {volume} {46}},\ \bibinfo {pages} {669--694} (\bibinfo {year}
		{1963})}\BibitemShut {NoStop}%
	\bibitem [{\citenamefont {Ma}(2006)}]{Ma06a}%
	\BibitemOpen
	\bibfield  {author} {\bibinfo {author} {\bibfnamefont {Zhong-Qi}\
			\bibnamefont {Ma}},\ }\bibfield  {title} {\enquote {\bibinfo {title} {The
				{L}evinson theorem},}\ }\href {http://stacks.iop.org/0305-4470/39/R625}
	{\bibfield  {journal} {\bibinfo  {journal} {Journal of Physics A:
				Mathematical and General}\ }\textbf {\bibinfo {volume} {39}},\ \bibinfo
		{pages} {R625--R659} (\bibinfo {year} {2006})}\BibitemShut {NoStop}%
	\bibitem [{\citenamefont {Dong}(2000)}]{Dong00b}%
	\BibitemOpen
	\bibfield  {author} {\bibinfo {author} {\bibfnamefont {Shi-Hai}\ \bibnamefont
			{Dong}},\ }\bibfield  {title} {\enquote {\bibinfo {title} {Levinson's theorem
				for the nonlocal interaction in one dimension},}\ }\href {\doibase
		10.1023/A:1003636110510} {\bibfield  {journal} {\bibinfo  {journal}
			{Int.~J.~Theor.~Phys.}\ }\textbf {\bibinfo {volume} {39}},\ \bibinfo {pages}
		{1529--1541} (\bibinfo {year} {2000})}\BibitemShut {NoStop}%
	\bibitem [{SLa()}]{SLaSi}%
	\BibitemOpen
	\href {http://slasi.knu.ua} {\enquote {\bibinfo {title} {\textsf{SLaSi}
				spin--lattice simulations package},}\ }\bibinfo {howpublished}
	{\url{http://slasi.knu.ua}},\ \bibinfo {note} {accessed:
		2016-04-17}\BibitemShut {NoStop}%
	\bibitem [{\citenamefont {Pylypovskyi}\ \emph {et~al.}(2014)\citenamefont
		{Pylypovskyi}, \citenamefont {Sheka}, \citenamefont {Kravchuk},\ and\
		\citenamefont {Gaididei}}]{Pylypovskyi14}%
	\BibitemOpen
	\bibfield  {author} {\bibinfo {author} {\bibfnamefont {Oleksandr~V.}\
			\bibnamefont {Pylypovskyi}}, \bibinfo {author} {\bibfnamefont {Denis~D.}\
			\bibnamefont {Sheka}}, \bibinfo {author} {\bibfnamefont {Volodymyr~P.}\
			\bibnamefont {Kravchuk}}, \ and\ \bibinfo {author} {\bibfnamefont {Yuri}\
			\bibnamefont {Gaididei}},\ }\bibfield  {title} {\enquote {\bibinfo {title}
			{Effects of surface anisotropy on magnetic vortex core},}\ }\href {\doibase
		http://dx.doi.org/10.1016/j.jmmm.2014.02.094} {\bibfield  {journal} {\bibinfo
			{journal} {Journal of Magnetism and Magnetic Materials}\ }\textbf {\bibinfo
			{volume} {361}},\ \bibinfo {pages} {201--205} (\bibinfo {year}
		{2014})}\BibitemShut {NoStop}%
	\bibitem [{Bay()}]{Bayreuth_cluster}%
	\BibitemOpen
	\href {http://www.hpc.uni-bayreuth.de/} {\enquote {\bibinfo {title} {High
				performance computing group at the {IT-ServiceCenter} of the {University of
					Bayreuth}},}\ }\bibinfo {howpublished}
	{\url{http://www.hpc.uni-bayreuth.de/}}\BibitemShut {NoStop}%
	\bibitem [{uni()}]{unicc}%
	\BibitemOpen
	\href {http://cluster.univ.kiev.ua/eng/} {\enquote {\bibinfo {title}
			{High--performance computing cluster of {T}aras {S}hevchenko {N}ational
				{U}niversity of {K}yiv},}\ }\bibinfo {howpublished}
	{\url{http://cluster.univ.kiev.ua/eng/}}\BibitemShut {NoStop}%
	\bibitem [{bit()}]{bitpcluster}%
	\BibitemOpen
	\href {http://horst-7.bitp.kiev.ua} {\enquote {\bibinfo {title} {Computing
				grid-cluster of the {B}ogolyubov {I}nsitute for {T}heoretical {P}hysics of
				{NAS} of {U}kraine},}\ }\bibinfo {howpublished}
	{\url{http://horst-7.bitp.kiev.ua}}\BibitemShut {NoStop}%
	\bibitem [{hyp()}]{hypnos}%
	\BibitemOpen
	\href {http://www.hzdr.de} {\enquote {\bibinfo {title} {{High Performance
					Computing at Helmholtz--Zentrum Dresden--Rossendorf}},}\ }\bibinfo
	{howpublished} {\url{http://www.hzdr.de}}\BibitemShut {NoStop}%
	\bibitem [{\citenamefont {Olver}\ \emph {et~al.}(2010)\citenamefont {Olver},
		\citenamefont {Lozier}, \citenamefont {Boisvert},\ and\ \citenamefont
		{Clark}}]{NIST10}%
	\BibitemOpen
	\bibinfo {editor} {\bibfnamefont {F.~W.~J.}\ \bibnamefont {Olver}}, \bibinfo
	{editor} {\bibfnamefont {D.~W.}\ \bibnamefont {Lozier}}, \bibinfo {editor}
	{\bibfnamefont {R.~F.}\ \bibnamefont {Boisvert}}, \ and\ \bibinfo {editor}
	{\bibfnamefont {C.~W.}\ \bibnamefont {Clark}},\ eds.,\ \href
	{http://www.cambridge.org/us/academic/subjects/mathematics/abstract-analysis/nist-handbook-mathematical-functions}
	{\emph {\bibinfo {title} {NIST Handbook of Mathematical Functions}}}\
	(\bibinfo  {publisher} {Cambridge University Press},\ \bibinfo {address} {New
		York, NY},\ \bibinfo {year} {2010})\BibitemShut {NoStop}%
	\bibitem [{\citenamefont {Griffiths}(1993)}]{Griffiths93}%
	\BibitemOpen
	\bibfield  {author} {\bibinfo {author} {\bibfnamefont {D~J}\ \bibnamefont
			{Griffiths}},\ }\bibfield  {title} {\enquote {\bibinfo {title} {Boundary
				conditions at the derivative of a delta function},}\ }\href {\doibase
		10.1088/0305-4470/26/9/021} {\bibfield  {journal} {\bibinfo  {journal}
			{Journal of Physics A: Mathematical and General}\ }\textbf {\bibinfo {volume}
			{26}},\ \bibinfo {pages} {2265--2267} (\bibinfo {year} {1993})}\BibitemShut
	{NoStop}%
\end{thebibliography}
%
%

%
\end{document}